\documentclass[aps,prb,reprint,amsmath,amssymb,longbibliography]{revtex4-2}
\usepackage{amsthm,mathtools}
\usepackage{microtype}
\usepackage{booktabs,array}
\usepackage[colorlinks,bookmarks=true,citecolor=blue,linkcolor=red,urlcolor=blue]{hyperref}
\usepackage{bookmark}
\hypersetup{
 pdfauthor={Glenn Wagner},
 pdftitle={The Halperin (111) state as a paired composite-fermion wave function},
 pdfsubject={An exact finite-size identity on the sphere}}

\theoremstyle{remark}

\begin{document}
\title{The Halperin (111) state as a paired composite-fermion wave function: \\ Two exact finite-size identities on the sphere}
\author{Glenn Wagner}
\affiliation{Institute for Theoretical Physics, ETH Z\"urich, 8093 Z\"urich, Switzerland}

\begin{abstract}
Two wavefunctions have previously been proposed to describe the $\nu=1/2+1/2$ quantum Hall bilayer: $p$-wave pairing of composite fermions and $s$-wave pairing of composite fermions with anti-composite fermions. We show analytically in the spherical geometry and within the Jain-Kamilla lowest Landau level projection scheme that both of these paired composite fermion wavefunctions \emph{exactly} reproduce the Halperin (111) state for a specific form of the pairing that we derive. We also find a second wavefunction that is common between both classes of wavefunctions. However, we provide evidence the two variational families generate overlapping but not identical sets of electronic states, ruling out a universal reparametrization that identifies their wavefunctions.
\end{abstract}

\maketitle

\section{Introduction}

The quantum Hall bilayer consists of two layers of a two-dimensional electron gas with layer densities $n_{\uparrow,\downarrow}$ in a large magnetic field $B$, separated by the interlayer spacing $d$. Defining the magnetic length as $\ell_b=\sqrt{\hbar/(eB)}$, the filling factors of the layers are $\nu_{\uparrow,\downarrow}=2\pi\ell_b^2n_{\uparrow,\downarrow}$.

The $\nu=1/2+1/2$ quantum Hall bilayer has a long history of experimental \cite{Experiment_Review} and theoretical \cite{Moon_Review,Eisenstein2004,Bonesteel,p_wave,Pwave2,Ezawa_2009,ShouCheng,HF1,HF2,HF3,HF4,HF5,HF6,ED1,ED3,ED0,DMRG,Park1,Park2,Simon1,Simon2,Simon3,Kimchi,Sodemann,Milovanovic,Ye1,Ye2,ICCFL,Cipri_thesis,CipriBonesteel,papicThesis,Bosonization1,Bosonization4,KunYang} studies. At large interlayer separations $d\gg\ell_b$, the two layers can be thought of as being decoupled. A single $\nu=1/2$ lowest Landau level is best described in terms of composite fermions (CFs), since they experience no net magnetic field at $\nu=1/2$ and therefore form a composite Fermi liquid \cite{CompositeFermionsJain,Jain2,HLR}. As $d/\ell_b$ is reduced, gauge fluctuations can mediate pairing \cite{Bonesteel,Kim2001bilayer,Deng2024pairing} and these composite fermions can form a paired state. In Refs.~\cite{Simon1,Simon2} a BCS state for the paired composite fermions was constructed and it was shown that the $p$-wave pairing channel is favourable. The $p$-wave channel was also shown to be favourable in a later Eliashberg calculation \cite{p_wave}. Using a variational wavefunction based on $p$-wave BCS paired CFs ($\Psi_{\textrm{BCS},p}$), it was shown that $\Psi_{\textrm{BCS},p}$ has excellent overlap with the exact diagonalization ground state on the sphere in the region $d/\ell_b\gtrsim1$ \cite{Simon1,Simon2}. 

On the other hand, when $d/\ell_b=0$ the system is described by the Halperin (111) wavefunction $\Psi_{111}$ \cite{Halperin111} which can be alternatively thought of as an exciton condensate of the electrons or as a quantum Hall ferromagnet \cite{Ezawa_2009,Moon_Review}. The relationship between the large $d/\ell_b$ and the small $d/\ell_b$ regime was a longstanding question.  

Recently, progress has been made on this question. The $p$-wave paired state of composite fermions was shown to be in the same phase as the Halperin (111) state \cite{Sodemann}. Furthermore, it was shown in \cite{Sodemann} that certain paired CF wavefunctions (which are different to the wavefunctions we are considering) can be related to the Halperin (111) state. This is supported by new exact diagonalization calculations which find good overlaps of the $p$-wave wavefunction with the exact diagonalization ground state for all $d/\ell_b$ \cite{Wagner2021,Hu2024single}. 

However, to obtain an intuitive understanding of the relationship between composite fermion wavefunctions and $\Psi_{111}$ it is more helpful to perform a particle-hole transformation on one of the layers and think of composite fermions in one layer pairing with anti-CFs in the other layer. In the limit where the CFs and anti-CFs are tightly bound, the fluxes attached should cancel and the state should describe an exciton condensate \cite{Crossover,Halperin2020}. Thus, as the interlayer separation is tuned, the system can be thought of as undergoing a BEC-BCS crossover \cite{Halperin2020,Sodemann,Wagner2021}, which is also backed up by experiments on double layer graphene \cite{Crossover}. A variational wavefunction based on this idea was constructed in which case the CFs and anti-CFs are paired in the $s$-wave channel, $\Psi_{\textrm{BCS},s}$. This wavefunction was shown to have very good overlaps with the exact diagonalization ground state for all $d/\ell_b$ \cite{Wagner2021,Hu2024single}. The $s$-wave channel was also shown to be favourable in an Eliashberg calculation \cite{Ruegg2023pairing,Lotric2024chern}. 

Thus, we have two different variational wavefunctions, $|\Psi_{\textrm{BCS},p}(g)\rangle$ and $|\Psi_{\textrm{BCS},s}(g)\rangle$, where $g$ is a set of variational parameters. The relationship between them is a puzzling question that has recently attracted interest. A field theoretic web of dualities argued that these states are in the same phase \cite{Ruegg2024dualities}. Furthermore, in Eliashberg theory the $s$-wave CF and anti-CF pairing has almost exactly the same pairing strength as the $p$-wave CF pairing \cite{Lotric2024chern}. Finally, numerically, both wavefunctions appeared to have extremely high overlaps with the Halperin (111) state \cite{Wagner2021,Hu2024single}. We will prove analytically and in the spherical geometry that for choices $g_p$ and $g_s$ which we provide, we have the \emph{exact} identities  $|\Psi_{\textrm{BCS},p}(g_p)\rangle=|\Psi_{111}\rangle$ and $|\Psi_{\textrm{BCS},s}(g_s)\rangle=|\Psi_{111}\rangle$. Thus the Halperin (111) wavefunction exactly lies in the variational space spanned by the paired composite fermion wavefunctions. 

The outline of the paper is as follows: In Sec.~\ref{sec:identities} we provide a statement of the two identities which have been proven. We also identify a second state shared between the two classes of wavefunctions.  Sec.~\ref{sec:s-proof} and Sec.~\ref{sec:p-proof} provide outlines of the proofs of the identities for $\Psi_{\textrm{BCS},s}$ and $\Psi_{\textrm{BCS},p}$ respectively. The supporting numerical evidence is left to App.~\ref{app:numerical} and the technical details of the proofs for $\Psi_{\textrm{BCS},s}$ and $\Psi_{\textrm{BCS},p}$ are given in Apps.~\ref{sec:supp-s} and \ref{sec:supp-p} respectively. Finally, we conclude in Sec.~\ref{sec:conclusion}, where we show that the two classes of wavefunctions do not cover identical regions of the Hilbert space.

\section{Statement of the identities}
\label{sec:identities}

In this section, we state the identities that we will prove. This section also serves to introduce the notation, especially concerning quantum Hall states in the spherical geometry \cite{Haldane1983fractional}. Further details of the notation and explicit expressions for all wavefunctions are provided in App.~\ref{sec:supp-orbitals}.

We consider a balanced bilayer in the spherical geometry with $N_1$ electrons per layer,
$N_e=2N_1$, and flux $N_\phi=2N_1-1\equiv2q$. All electrons
are confined to the lowest Landau level (LLL) and the physical spin is assumed to be polarized. Overall
coordinate-independent factors are suppressed throughout.

Let $X=(x_1,\ldots,x_{N_1})$ and $Y=(y_1,\ldots,y_{N_1})$
denote the electron coordinates in the two layers. For a point
$x_i=(\theta_i,\phi_i)$ on the sphere, define the spinor coordinates
\begin{equation}
    u_i=\cos\frac{\theta_i}{2}e^{-i\phi_i/2},
    \qquad
    v_i=\sin\frac{\theta_i}{2}e^{i\phi_i/2},
    \label{eq:spinors}
\end{equation}
and $[x_i,x_j]\equiv u_iv_j-v_iu_j$. For a coordinate set
$Z=(z_1,\ldots,z_{N_1})$, the Jastrow factors are
\begin{equation}
    J(Z)=\prod_{i<j}[z_i,z_j],
    \qquad
    J_i(Z)=\prod_{j\ne i}[z_i,z_j].
    \label{eq:jastrows}
\end{equation}

The single-particle eigenstates of electrons on a sphere penetrated by flux are given by the monopole harmonics  \cite{Identities,WU1976365}. We denote the orthonormal monopole harmonics at flux $2q$
by $Y_{q,n,m}$, where $n=0,1,\ldots$ is the Landau-level
index, $\ell=q+n$ is the angular momentum, and
$m=-\ell,-\ell+1,\ldots,\ell$. We use the spinor gauge in
which $Y_{q,0,m}(x)\propto u(x)^{q+m}v(x)^{q-m}$.

The Halperin (111) state \cite{Halperin111} in electron coordinates is
\begin{equation}
    \Phi_{111}(X,Y)
    =J(X)J(Y)\prod_{i,j=1}^{N_1}[x_i,y_j].
    \label{eq:111-polynomial}
\end{equation}
After a particle--hole transformation in the lower layer,
there are $N_1$ holes, with coordinates
$H=(h_1,\ldots,h_{N_1})$. Using the complex-conjugated
LLL orbitals for holes, the same state becomes
\begin{equation}
    \Phi_{111}^{\mathrm{ph}}(X,H)
    =
    \det_{i,j}\!\left[
        \sum_{m=-q}^{q}
        Y_{q,0,m}(x_i)Y_{q,0,m}(h_j)^*
    \right].
    \label{eq:111-ph}
\end{equation}
To see this, it is easiest to perform the particle-hole transformation using the second quantized form of the Halperin (111) state and then transform back to real space at the end. The corresponding version in the disk geometry was derived in \cite{KunYang}.

Attaching two flux quanta within each layer leaves the
effective CF monopole strength $Q=q-(N_1-1)=1/2$. The LLL projection is performed using the Jain-Kamilla procedure \cite{KamillaJain,Kamilla_thesis,JainKamilla_IJMP}.
We denote the Jain--Kamilla projected counterpart of
$Y_{Q,n,m}(z_i)$ by
$\widetilde Y_{Q,n,m}(z_i;Z)$: the orbital is projected
together with its individual Jastrow factor $J_i(Z)$,
which is subsequently divided out. The resulting orbital
depends on all coordinates in $Z$. We abbreviate it as
$\widetilde Y_{nm;i}(Z)$. Its explicit differential form
and normalization conventions are specified in App.~\ref{app:JK}.
The nonvanishing projected shells are $n=0,\ldots,N_1-1$,
with angular momentum $\ell_n=n+1/2$.

For pairing coefficients
$\boldsymbol g=(g_0,\ldots,g_{N_1-1})$, define
\begin{align}
    (G_p)_{ij}
    &=
    \sum_{n=0}^{N_1-1}g_n
    \sum_{m=-\ell_n}^{\ell_n}
    (-1)^{m+1/2}
    \widetilde Y_{nm;i}(X)
    \widetilde Y_{n,-m;j}(Y),
    \label{eq:p-pairing-kernel}\\
    (G_s)_{ij}
    &=
    \sum_{n=0}^{N_1-1}g_n
    \sum_{m=-\ell_n}^{\ell_n}
    \widetilde Y_{nm;i}(X)
    \widetilde Y_{nm;j}(H)^*.
    \label{eq:s-pairing-kernel}
\end{align}
The complex-conjugated hole-layer orbitals describe anti-CFs.
Multiplying the CF wavefunctions by Jastrow factors describing the flux attachment, we obtain the corresponding paired wavefunctions as \cite{Simon1,Simon2,Wagner2021,Hu2024single}
\begin{align}
    \Phi_{\mathrm{BCS},p}(\boldsymbol g;X,Y)
    &=J(X)^2J(Y)^2\det G_p,
    \label{eq:p-wavefunction}\\
    \Phi_{\mathrm{BCS},s}^{\mathrm{ph}}(\boldsymbol g;X,H)
    &=J(X)^2[J(H)^*]^2\det G_s.
    \label{eq:s-wavefunction}
\end{align}
Our central result is that the coefficient choices
\begin{align}
    g_{p,n}^{(111)}
    &=
    (n+1)
    \frac{(N_1-1-n)!}{(N_1+n+1)!}
    \left[\frac{(2N_1+n)!}{(2N_1)!}\right]^2,
    \label{eq:exact-p-coefficients}\\
    g_{s,n}^{(111)}
    &=\delta_{n,N_1-1},
    \label{eq:exact-s-coefficients}
\end{align}
give the exact identities
\begin{align}
    \Phi_{\mathrm{BCS},p}(\boldsymbol g_p^{(111)};X,Y)
    &\propto\Phi_{111}(X,Y),
    \label{eq:exact-p-identity}\\
    \Phi_{\mathrm{BCS},s}^{\mathrm{ph}}(\boldsymbol g_s^{(111)};X,H)
    &\propto\Phi_{111}^{\mathrm{ph}}(X,H).
    \label{eq:exact-s-identity}
\end{align}
Thus, both paired CF constructions
reproduce the (111) electron state exactly at finite $N_1$. We provide the analytical expressions for the normalization of the wavefunctions in App.~\ref{app:normalization}. 

The $s$-wave and $p$-wave representations of the Halperin (111) are therefore very different: The $s$-wave case only has a single CF shell occupied, namely the shell with the largest $n$ at a given system size. On the other hand, the $p$-wave case has non-zero occupation of all CF shells. As we show in App.~\ref{app:p_wave_occupations}, in the $p$-wave case the occupation probability of each shell decreases monotonically with $n$. 

By combining a particle--hole transformation in the lower
layer with the antipodal substitution
$(u_h,v_h)=(-v_y^*,u_y^*)$, one can interchange the two constructions, while sending $g_n\mapsto(-1)^n g_n$, up to an overall phase. Thus the state with
$g_{p,n}=g_{s,n}^{(111)}$ and
$g_{s,n}=(-1)^n g_{p,n}^{(111)}$ is also shared between both variational states. The electron-pair amplitude is maximal at antipodal points, whose
separation diverges with the sphere radius; this additional common
state therefore does not describe local interlayer pairing in the
thermodynamic limit. We provide further details of this state in App.~\ref{app:yang-state}.

\section{Proof of equality of 111 state and $s$-wave paired CFs and anti-CFs}
\label{sec:s-proof}

In this section we will outline the proof of Eq.~\eqref{eq:exact-s-identity}, with detailed algebraic manipulations given in App.~\ref{sec:supp-s}. As stated in Eq.~\eqref{eq:exact-s-coefficients}, the suitable choice of variational parameters is $g_{s,n}^{(111)}=\delta_{n,N_1-1}$. Numerically, this choice was already observed to reproduce the Halperin (111) state in Ref.~\cite{Hu2024single}. For this choice of $g_{s,n}$ only the highest CF shell
contributes.  Its angular momentum is
$\ell_{N_1-1}=q$, so it contains the same $2N_1$ magnetic
orbitals as the physical LLL. The proof relates the Slater
determinants formed from these two sets of orbitals.

Relabel the magnetic quantum number by
$r=q-m=0,\ldots,N_\phi$ and, for any layer coordinates $Z$,
define the $N_1\times 2N_1$ matrices
\begin{align}
    F_{ir}(Z)
    &=J_i(Z)\,
      \widetilde Y_{N_1-1,\,q-r;i}(Z),
    \label{eq:s-F-matrix}\\
    V_{ir}(Z)
    &=Y_{q,0,q-r}(z_i).
    \label{eq:s-V-matrix}
\end{align}
With these definitions, the paired state Eq.~\eqref{eq:s-wavefunction} becomes
\begin{equation}
    \Phi_{\mathrm{BCS},s}^{\mathrm{ph}}
    (\boldsymbol g_s;X,H)
    =\det\!\left[F(X)F(H)^\dagger\right].
    \label{eq:s-gram-determinant}
\end{equation}
Similarly, Eq.~\eqref{eq:111-ph} gives
\begin{equation}
    \Phi_{111}^{\mathrm{ph}}(X,H)
    =\det\!\left[V(X)V(H)^\dagger\right].
    \label{eq:s-111-gram-determinant}
\end{equation}
We will now derive a relation between $F$ and $V$. 
At the highest shell, the derivatives in the definition of $\widetilde Y$ exhaust the degree of the individual Jastrow
factor. 
We obtain
\begin{align}
F_{ir}(Z)\propto{}&\nonumber
\frac{(-1)^r}{\sqrt{\binom{N_\phi}{r}}}
\int_0^{2\pi}\frac{d\phi}{2\pi}\,
e^{-im\phi}\left(u_i e^{i\phi/2}+v_i e^{-i\phi/2}\right)^{N_1-1}
\\[-2pt]
&\times
\prod_{k=1}^{N_1}
\left(u_k e^{i\phi/2}+v_k e^{-i\phi/2}\right),
\label{eq:s-generating-orbitals}
\end{align}
where the integral extracts the part with azimuthal angular momentum $m=q-r$. The
suppressed factor is independent of $i$, $r$, and $Z$.
The detailed reduction of the projected orbitals to
Eq.~\eqref{eq:s-generating-orbitals} is given in 
App.~\ref{app:generating}.

In App.~\ref{app:orthogonality} we show that summing over the magnetic quantum numbers gives the bilinear orthogonality relation
\begin{align}
    (F(Z)CV(Z)^T)_{ij}\nonumber
\propto&
(u_i v_j-v_i u_j)^{N_1-1}\\
&\times\prod_{k=1}^{N_1}(u_kv_j-v_ku_j)
=0,
\end{align}
where the expression vanishes since it contains the term $k=j$. Here we defined the matrix $C_{rs}=(-1)^r\delta_{r+s,N_\phi}$.
For generic coordinates, $F$ and $W=VC^T$ both have rank $N_1$
and satisfy $FW^T=0$. Row operations, with a common relabelling
of columns if necessary, reduce them to $F'=(I,T)$ and
$W'=(-T^T,I)$. Consider
\[
  M=\begin{pmatrix}I&T\\0&I\end{pmatrix},
  \qquad
  M^{-1}=\begin{pmatrix}I&-T\\0&I\end{pmatrix}.
\]
The first $N_1$ rows of $M$ are $F'$, while the last $N_1$
columns of $M^{-1}$ are $(W')^T$. Jacobi's complementary-minor
identity~\cite{BrualdiSchneider1983} is:
\begin{equation}
  \det M_{A,R}
  =
  \pm\,\det M\,
  \det(M^{-1})_{R^c,A^c},
\end{equation}
where $M_{A,R}$ denotes the submatrix with row set $A$ and column set $R$, and complements are taken within $\{0,\ldots,2N_1-1\}$. Choosing $A=\{0,\ldots,N_1-1\}$ and using $\det M=1$
therefore relates a minor
of $F'$ on columns $R$ to a minor of $W'$ on the complementary
columns $R^c$. Undoing the row operations introduces only a
common factor. Finally, the antidiagonal matrix $C$ reverses
the column labels, giving the reversed complement 
\begin{equation}
    \rho(R^c)=\{N_\phi-r:\ r\notin R\}.
    \label{eq:s-reversed-complement}
\end{equation}
We now obtain the central result which is a relation between the maximal minors (the determinants of the largest possible square submatrices)
of $F$ and $V$. Let $R$ be an $N_1$-element subset of
$\{0,\ldots,N_\phi\}$, and let $F_R$ and $V_R$ denote the
submatrices containing those columns in increasing order, then
\begin{equation}
    \det F_R(Z)\propto
    \det V_{\rho(R^c)}(Z).
    \label{eq:s-complementary-minors}
\end{equation}

It remains to establish that the common proportionality
factor is independent of the coordinates. Choose $R_0=\{0,\ldots,N_1-1\}$, for which $\rho(R_0^c)=R_0$. On the patch where all $u_k\ne0$, homogeneity allows us to set $u_k=1$ without changing the ratio of the two minors. The components with $r=0,\ldots,N_1-1$ are then common triangular linear combinations of $1,v_i,\ldots,v_i^{N_1-1}$, with nonzero diagonal coefficients depending only on $N_1$. Eq.~\eqref{eq:s-generating-orbitals}
therefore reduces this minor to a Vandermonde determinant (see App.~\ref{app:evaluation}):
\begin{equation}
    \det F_{R_0}(Z)
    \propto
    \prod_{i<j}(v_j-v_i)
    \propto
    \det V_{R_0}(Z).
    \label{eq:s-vandermonde-minor}
\end{equation}
All factors removed in this step depend only on $N_1$.
Both minors are homogeneous of degree $N_\phi$ in each
spinor, so the result extends beyond this affine patch.
This establishes Eq.~\eqref{eq:s-complementary-minors}.

The identity now follows by applying the Cauchy-Binet equation to Eq.~\eqref{eq:s-gram-determinant}. The Cauchy-Binet equation allows the determinant of a product of rectangular matrices to be written as a sum over products of determinants of square submatrices \cite{BrualdiSchneider1983}: 
\begin{align}
    \Phi_{\mathrm{BCS},s}^{\mathrm{ph}}
    &=
    \sum_{|R|=N_1}
    \det F_R(X)\,\det F_R(H)^*
    \nonumber\\
    &\propto
    \sum_{|R|=N_1}
    \det V_{\rho(R^c)}(X)\,
    \det V_{\rho(R^c)}(H)^*
    \nonumber\\
    &=
    \det\!\left[V(X)V(H)^\dagger\right]
    =
    \Phi_{111}^{\mathrm{ph}}(X,H).
    \label{eq:s-proof-result}
\end{align}
Here $R\mapsto\rho(R^c)$ simply permutes the
$N_1$-element subsets.

\section{Proof of equality of 111 state and $p$-wave paired CFs}
\label{sec:p-proof}

In this section we will prove Eq.~\eqref{eq:exact-p-identity}, with detailed algebraic manipulations given in App.~\ref{sec:supp-p}. The
Halperin (111) state can be re-written using the Cauchy identity as \cite{Kim2001bilayer,Ho1995broken,Simon1}
\begin{equation}
     \Phi_{111}(X,Y)=
    \left(\prod_{i,j=1}^{N_1}[x_i,y_j]\right)^2
    \det_{i,j}\!\left[\frac{1}{[x_i,y_j]}\right].
    \label{eq:111-cauchy}
\end{equation}
For the $p$-wave state, we first reduce the projected
pairing kernel to a product of spinor brackets and show that this reproduces Eq.~\eqref{eq:111-cauchy}. Fix $\boldsymbol g=\boldsymbol g_p^{(111)}$ from
Eq.~\eqref{eq:exact-p-coefficients} and absorb the
individual Jastrow factors into the pairing matrix:
\begin{equation}
    K_{ij}=J_i(X)J_j(Y)(G_p)_{ij}.
    \label{eq:p-polynomial-kernel}
\end{equation}
Each entry is now a polynomial and
\begin{equation}
    \Phi_{\mathrm{BCS},p}(\boldsymbol g_p^{(111)};X,Y)
    =\det K.
    \label{eq:p-determinant}
\end{equation}
The central identity is the factorization
\begin{equation}
    K_{ij}\propto
    [x_i,y_j]
    \prod_{k\ne i}[x_k,y_j]
    \prod_{\ell\ne j}[x_i,y_\ell],
    \label{eq:p-kernel-factorization}
\end{equation}
with a nonzero proportionality factor depending only
on $N_1$. In particular, it is the same for every
matrix element.

We outline how the chosen pairing coefficients produce
this identity. The right hand side of Eq.~\eqref{eq:p-kernel-factorization} is a holomorphic polynomial,
homogeneous of degree $N_1$ in each distinguished
spinor $x_i,y_j$ and degree one in every spectator
spinor. The $m,-m$ contraction in Eq.~\eqref{eq:p-pairing-kernel} is a rotational singlet,
so both sides are invariant under rotation of all electron coordinates by the same amount on the sphere. 
A common rotation about the $z$ axis acts as $(u_k,v_k)\mapsto(e^{-i\theta/2}u_k,e^{i\theta/2}v_k)$. For a holomorphic polynomial to be invariant under these rotations, each monomial must contain equal total powers of $u$ and $v$, summed over all particles. It is therefore also invariant under the reciprocal rescaling $(u_k,v_k)\mapsto(\lambda u_k,\lambda^{-1}v_k)$ with $\lambda>0$. Every $L\in SL(2,\mathbb C)$ can be decomposed into two $SU(2)$ rotations with such a rescaling between them. Thus rotational invariance of a holomorphic polynomial implies invariance under the simultaneous action of $SL(2,\mathbb C)$ on all spinors~\cite{pakatchi2022quantumhallgroundstates,CAN2015752}. To apply this observation to the two sides of our kernel identity, define
\begin{align}
 S&=\begin{pmatrix}u_i&u_j\\v_i&v_j\end{pmatrix},
 \qquad w=\det S=[x_i,y_j],\nonumber\\
 L&=\begin{pmatrix}1&0\\0&w\end{pmatrix}S^{-1}
\end{align}
where $(u_i,v_i)$ and $(u_j,v_j)$ are the spinors of $x_i$ and $y_j$.
Applying the same $L$ to every spinor in both layers sends the distinguished spinors to $(1,0)$ and $(0,w)$ while leaving both polynomials unchanged. Keeping these spinors unnormalized, homogeneity of degree $N_1$ in the second distinguished spinor supplies a common factor $w^{N_1}$, thus it allows us to check the identity for
\begin{equation}
    \bigl(u_i,v_i\bigr)=(1,0),
    \qquad
    \bigl(u_j,v_j\bigr)=(0,1),
    \label{eq:p-poles}
\end{equation}
For this choice, Eq.~\eqref{eq:p-kernel-factorization} becomes
\begin{equation}
    K_{ij}
\propto
\prod_{k\ne i}u(x_k)\,
\prod_{\ell\ne j}v(y_\ell).
\end{equation}
Expand the individual Jastrow factors in their distinguished spinors:
\begin{align}
    J_i(X)
    &=
    \sum_{r=0}^{N_1-1}
    (-1)^r A_r u_i^{N_1-1-r}v_i^r,
    \label{eq:p-spectator-A}\\
    J_j(Y)
    &=
    \sum_{r=0}^{N_1-1}
    (-1)^{N_1-1-r}B_r u_j^r v_j^{N_1-1-r}.
    \label{eq:p-spectator-B}
\end{align}
where
\begin{equation}
    A_{N_1-1}=\prod_{k\ne i}u(x_k),
    \qquad
    B_{N_1-1}=\prod_{\ell\ne j}v(y_\ell).
    \label{eq:p-last-coefficients}
\end{equation}
We therefore want to choose $g_{p,n}^{(111)}$ to obtain $K_{ij}\propto A_{N_1-1}B_{N_1-1}$.
With the choice Eq.~\eqref{eq:p-poles} for the positions of the distinguished spinors, substitution of the projected orbitals $\widetilde Y$
and coefficients $g_{p,n}^{(111)}$ into Eq.~\eqref{eq:p-polynomial-kernel} gives (see App.~\ref{app:distinguished_poles})
\begin{equation}
    K_{ij}\propto
    \sum_{r=0}^{N_1-1}
    (-1)^r(N_1-1-r)!^2 A_rB_r\,S_r,
    \label{eq:p-pole-expansion}
\end{equation}
where
\begin{align}
    S_r&=
    \sum_{n=r}^{N_1-1}
    (-1)^n g_{p,n}^{(111)}
    \frac{(2N_1)!^2(n+r+1)!}
    {(2N_1+n)!^2(n-r)!(N_1-1-n)!^2}
    \nonumber\\
    &=\frac{1}{2}(-1)^{N_1-1}\,
    \delta_{r,N_1-1}.
    \label{eq:p-finite-sum}
\end{align}
See App.~\ref{app:telescoping} for the evaluation of the sum. Thus, for the particular choice of $g_{p,n}^{(111)}$, only $A_{N_1-1}B_{N_1-1}$ survives in
Eq.~\eqref{eq:p-pole-expansion}. This explains the choice of $g_{p,n}^{(111)}$. 

Covariance and homogeneity
then establish the factorization for arbitrary
linearly independent distinguished spinors.
Polynomiality extends it to the remaining
configurations (see App.~\ref{app:extension}).

To evaluate the determinant, rewrite the kernel Eq.~\eqref{eq:p-kernel-factorization} as
\begin{equation}
    K_{ij}\propto
    \left(\prod_{\ell=1}^{N_1}[x_i,y_\ell]\right)
    \frac{1}{[x_i,y_j]}
    \left(\prod_{k=1}^{N_1}[x_k,y_j]\right).
    \label{eq:p-cauchy-kernel}
\end{equation}
Extracting the first product from each row and the
last product from each column gives (see App.~\ref{app:Cauchy})
\begin{equation}
    \det K\propto
    \left(\prod_{i,j=1}^{N_1}[x_i,y_j]\right)^2
    \det_{i,j}\!\left[\frac{1}{[x_i,y_j]}\right],
    \label{eq:p-cauchy-reduction}
\end{equation}
which coincides with the form of the 111 state in Eq.~\eqref{eq:111-cauchy}.

\section{Conclusion}
\label{sec:conclusion}

We have shown analytically that both the $p$-wave paired
CF--CF wavefunction and the $s$-wave paired CF--anti-CF
wavefunction exactly reproduce the Halperin (111) state
for explicit choices of their pairing amplitudes. These
identities hold for arbitrary balanced particle number
on the sphere within the Jain--Kamilla projection scheme \cite{KamillaJain,Kamilla_thesis,JainKamilla_IJMP}.
For the $s$-wave construction, the equality is established
in electron--hole coordinates.

These results provide an explicit microscopic connection
between interlayer exciton condensation and the two paired
composite-fermion descriptions. At the identified parameter
choices, the different pairing channels encode identical
electronic correlations. The explicit pairing amplitudes
also provide exact benchmarks for numerical calculations
and starting points for variational optimization at
nonzero interlayer separation.

A central question is how the variational spaces
spanned by the two families compare as their pairing
amplitudes are varied. With both constructions expressed
in electron coordinates, our identities establish that
their intersection contains the (111) state. It remains
to determine what the total intersection of the two spaces spanned by the two wavefunctions is. One can first ask this question in the perturbative sense, for small variations away from the (111) state, how do the two wavefunction families evolve? However, one may ask in a more global sense whether all states are shared between the two variational wavefunctions? 

There is evidence against all states being shared. Firstly, in Ref.~\cite{Wagner2021}, the overlaps of the exact diagonalization state at finite $d/\ell_b$ with the two different optimized variational wavefunctions are similar but not identical. Secondly, at large interlayer separation, they tend to different limits, while the $p$-wave paired state describes two decoupled composite Fermi liquids in the two layers, the $s$-wave paired state describes a composite Fermi liquid in one layer and an anti-composite Fermi liquid in the other layer. The composite Fermi liquid and anti-composite Fermi liquid do not coincide in the standard Halperin-Lee-Read construction \cite{HLR,AntiCFLiquid,Son,Balram2016nature,Chong2017particle}, although the wavefunction overlaps are extremely close to one \cite{RezayiHaldane}. Finally, in App.~\ref{app:sp-counterexample} we construct a state of the $s$-wave family for $N_1=4$ that cannot be represented by the $p$-wave family, thus providing a counter-example showing that the two variational families do not coincide at $N_1=4$.

In this work we considered the case of $\nu=1/2+1/2$ quantum Hall bilayers. A natural extension would be to consider the case of imbalanced quantum Hall bilayers such as $\nu=1/4+3/4$ and $\nu=1/3+2/3$ where similar wavefunctions of paired composite fermions have been used \cite{Wagner2024,Wagner2025variational}. Another natural extension would be to consider composite fermion wavefunctions for more than two layers, where generalized Halperin wavefunctions have been proposed \cite{Goerbig2007analysis,Wang2026interlayer,das2026threedimensionalfoliatedfractionalquantum}.

Beyond quantum Hall bilayers, Halperin states can also be used to describe two-component states where both components have opposite Chern number \cite{Sodemann2024Halperin,das2025fractionaltopologicalinsulatorsoddinteger,Brunner2026tuning}. To do so, a particle-hole transformation in one spin species is performed, as was done for our $s$-wave pairing analysis. These Halperin states may be relevant in the study of moiré materials such as twisted MoTe$_2$. Twisted MoTe$_2$ can host a zero-field composite Fermi liquid \cite{Goldman2023cfl,Dong2023cfl,aguilarméndez2026fractionalcherninsulatorscomposite}, and the CF perspective may shed light on the energetic stabilization of the putative partially particle-hole transformed Halperin states.

\begin{acknowledgments}
G.W.~acknowledges support from the Swiss National Science Foundation (SNSF)
through Ambizione Grant No.~PZ00P2-216183.
ChatGPT was used to assist with developing the proof and preparing the
manuscript.
\end{acknowledgments}

\bibliography{bib}

\clearpage
\newpage

\begin{appendix}
\onecolumngrid
	\begin{center}
		\textbf{\large --- Supplementary Material ---\\The Halperin (111) state as a paired composite-fermion wave function: \\ Two exact finite-size identities on the sphere}\\
		\medskip
		\text{Glenn Wagner}
	\end{center}

		\setcounter{equation}{0}
	\setcounter{figure}{0}
	\setcounter{table}{0}
	\makeatletter
    \@removefromreset{equation}{section}
    \makeatother
	\renewcommand{\theequation}{S\arabic{equation}}
	\renewcommand{\thefigure}{S\arabic{figure}}
    \renewcommand{\thetable}{S\arabic{table}}
	\renewcommand{\bibnumfmt}[1]{[S#1]}

\section{Numerical checks}
\label{app:numerical}

\subsection{Default precision arithmetic}

\begin{figure}[h]
    \centering
    \includegraphics[width=0.7\linewidth]{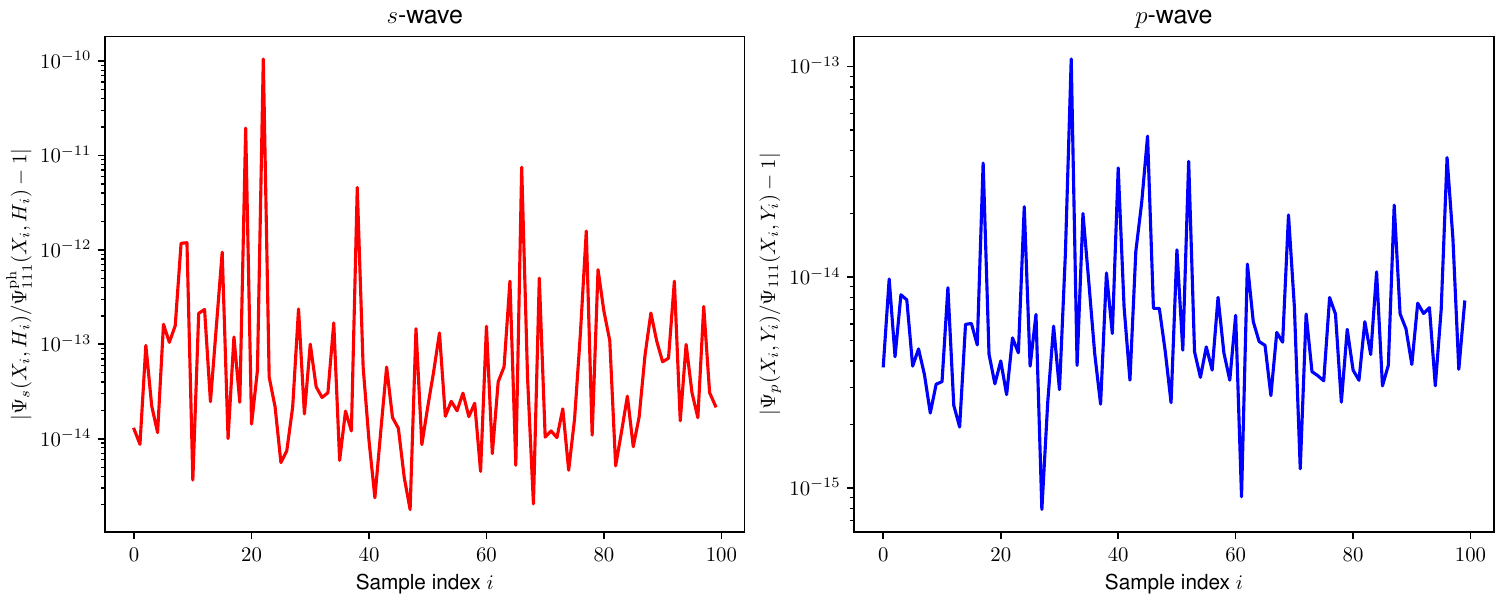}
    \caption{\textbf{Relative residuals of wavefunctions for 4+4 electrons.} We plot the absolute deviation of the normalized wavefunction ratios from unity of the two composite fermion wavefunctions to the Halperin (111) wavefunction for 100 randomly chosen configurations of the particles.  }
    \label{fig:overlap_s_p_111}
\end{figure}

\begin{figure}[h]
    \centering
    \includegraphics[width=0.7\linewidth]{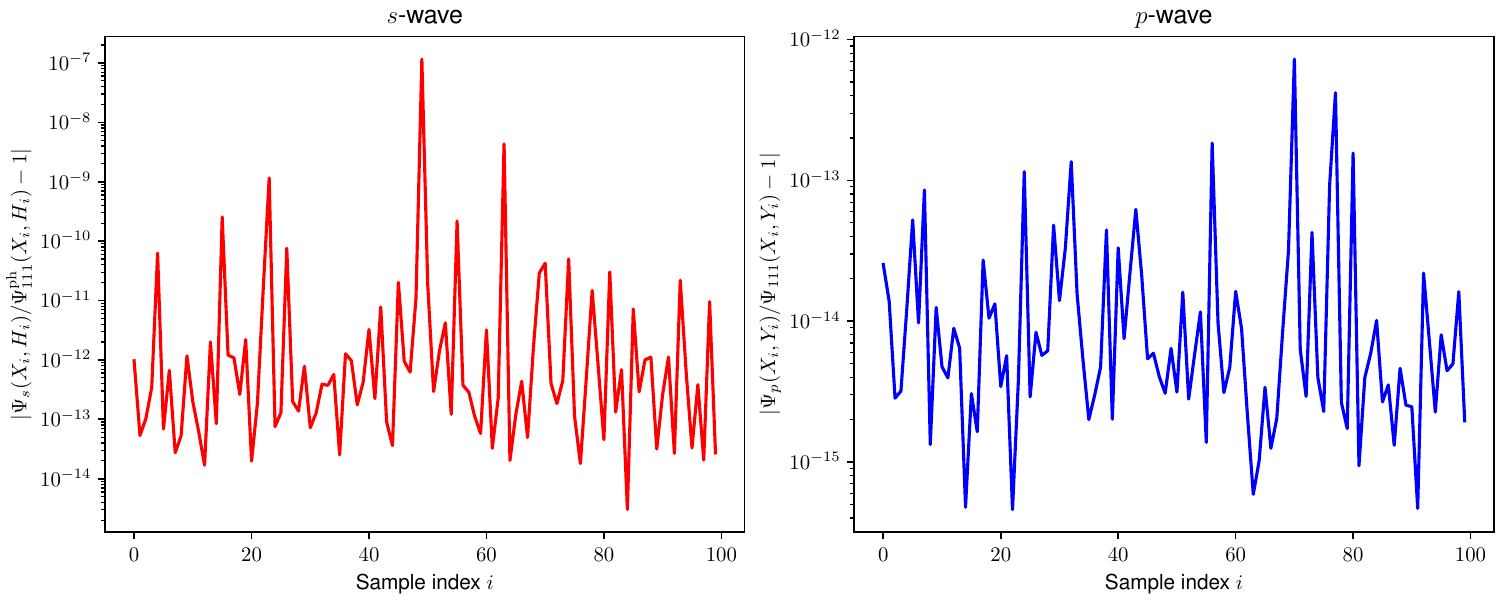}
    \caption{\textbf{Relative residuals of wavefunctions for 5+5 electrons.} We plot the absolute deviation of the normalized wavefunction ratios from unity of the two composite fermion wavefunctions to the Halperin (111) wavefunction for 100 randomly chosen configurations of the particles.  }
    \label{fig:overlap_s_p_111_Ne_10}
\end{figure}

\begin{figure}[h]
    \centering
    \includegraphics[width=0.7\linewidth]{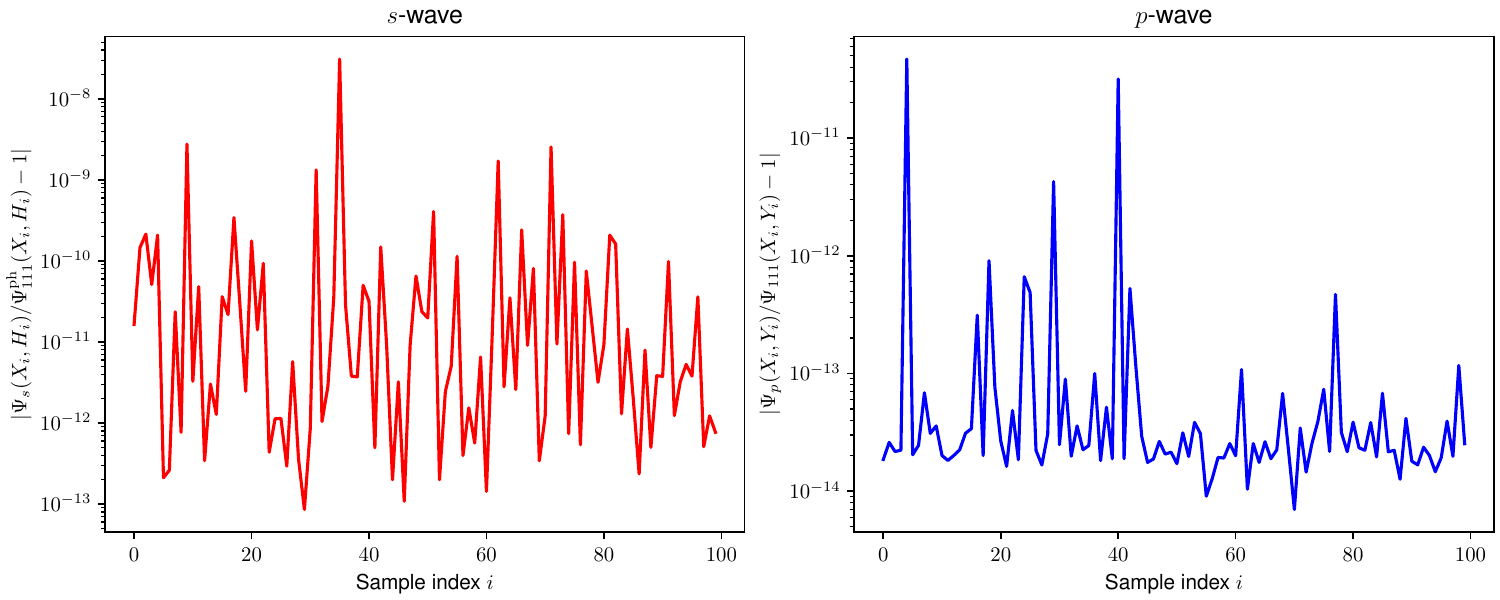}
    \caption{\textbf{Relative residuals of wavefunctions for 6+6 electrons.} We plot the absolute deviation of the normalized wavefunction ratios from unity of the two composite fermion wavefunctions to the Halperin (111) wavefunction for 100 randomly chosen configurations of the particles.  }
    \label{fig:overlap_s_p_111_Ne_12}
\end{figure}

In Figs.~\ref{fig:overlap_s_p_111}, \ref{fig:overlap_s_p_111_Ne_10}, \ref{fig:overlap_s_p_111_Ne_12} we confirm the proven identity numerically for $4+4$, $5+5$ and $6+6$ electrons respectively. For each system size, we sample 100 particle configurations, drawing the angles $\theta\in[0,\pi]$ and $\phi\in[0,2\pi)$ of each electron or hole independently and uniformly. The agreement is better for smaller system sizes and it is better for the $p$-wave case than for the $s$-wave case. We attribute deviations from floating point accuracy to finite precision arithmetic when evaluating the wavefunctions, e.g.~overflow/underflow issues especially when computing the determinant of small numbers. For the $p$-wave case there is significant weight in the lower shells, which are less complex to evaluate, while for the $s$-wave case all CFs are in the highest shell; this explains why the $p$-wave case is closer to floating point accuracy. Larger system sizes also are more complex which reduces the accuracy. Here we used the default \texttt{float64} and \texttt{complex128} types in Python. 

\subsection{High precision arithmetic}

\begin{figure}[h]
    \centering
    \includegraphics[width=0.8\linewidth]{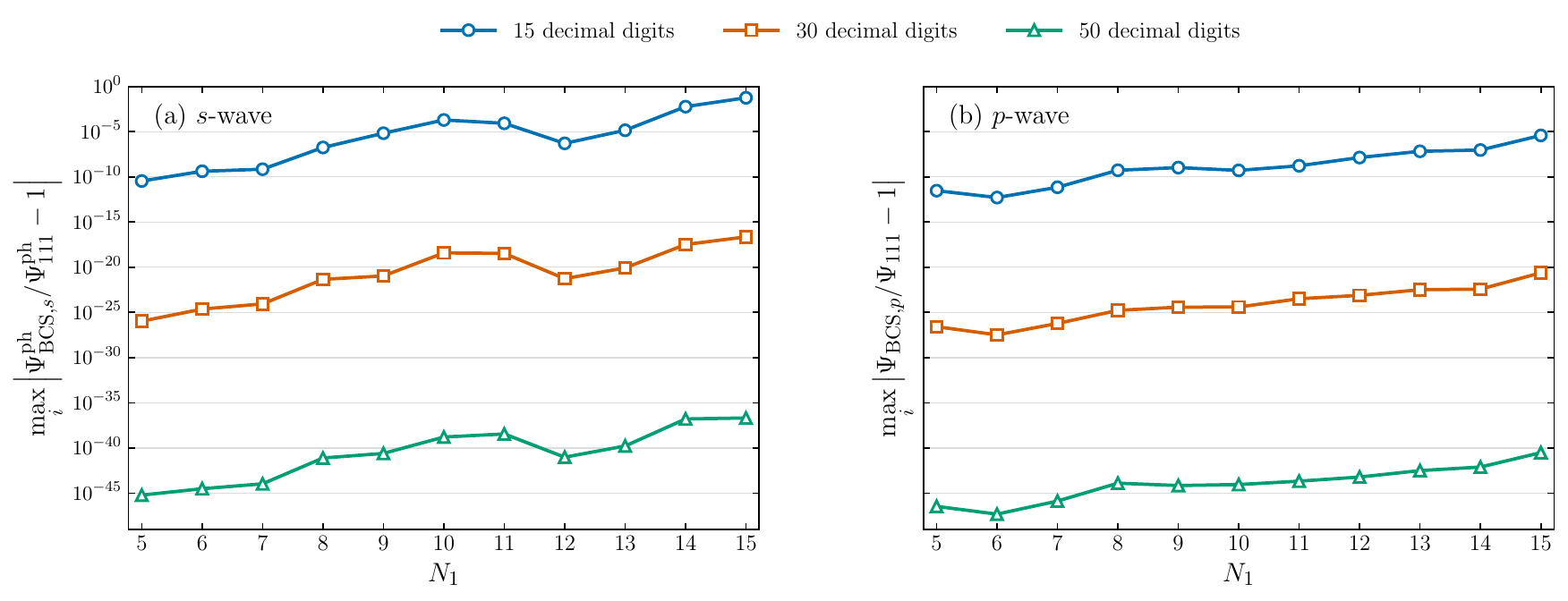}
    \caption{\textbf{Relative residuals of wavefunctions using high precision arithmetic.} Maximum relative residual between (a) the $s$-wave CF--BCS state $\Psi_{\mathrm{BCS},s}^{\mathrm{ph}}$ and the particle--hole-transformed $(111)$ state $\Psi_{111}^{\mathrm{ph}}$, and (b) the $p$-wave CF--BCS state $\Psi_{\mathrm{BCS},p}$ and the $(111)$ state $\Psi_{111}$, for balanced systems with $N_1=5,\ldots,15$. For each system size, the maximum is taken over $100$ random coordinate configurations, with the polar and azimuthal angles sampled independently and uniformly. The same configurations are evaluated using arbitrary-precision arithmetic with $15$, $30$, and $50$ decimal digits. The wavefunctions are compared with matched overall normalization and phase conventions. Increasing the arithmetic precision suppresses the residuals in both representations, consistent with the exact analytical identities.}
    \label{fig:high_precision}
\end{figure}

To obtain better accuracy, high precision arithmetic has to be used, as was done in Ref.~\cite{Davenport2012spinful}. Here, we use the \texttt{mpmath} Python package to fix the working precision. As shown in Fig.~\ref{fig:high_precision}, the agreement between the different wavefunctions improves as the decimal precision is improved. The number of decimal places that need to be kept increases as the system size $N_1$ increases. We note that the loss of precision is the reason why in Ref.~\cite{Hu2024single} the overlap of the $s$-wave paired CF wavefunction with the Halperin (111) wavefunction was found to be small at the largest system size used ($10+10$), but extremely close to one for all smaller system sizes. Finally, we note that even better numerical accuracy and faster evaluations may also be obtained using the recently developed quaternion method for JK projection \cite{Anakru2025emergent,Gattu2025unlocking}. While the usual JK projection can be used for up to 100 electrons \cite{adlong2026anyonsimpuritiesfractionalcharge}, the more recent methods allow for significantly larger system sizes to be studied.

\section{Monopole harmonics and projected-orbital conventions}
\label{sec:supp-orbitals}

We retain the notation of the main text:
$N_e=2N_1$, $N_\phi=2N_1-1=2q$, and the effective CF
monopole strength is $Q=1/2$. The spinors are
$u=\cos(\theta/2)e^{-i\phi/2}$ and
$v=\sin(\theta/2)e^{i\phi/2}$, with
$[x,y]=u(x)v(y)-v(x)u(y)$.
All derivatives below act on the distinguished particle spinor,
holding every spectator spinor fixed. Unlike the main text,
we keep coordinate-independent factors and specify all orbital phases.

\subsection{Harmonics and the one-coordinate projection rule}

For a nonnegative monopole strength $Q$, we choose the harmonics \cite{WU1976365,Identities}
\begin{align}
 Y_{Q,n,m}(u,v)
 ={}&\mathcal N_{Qnm}(-1)^{Q+n-m}
 \sum_{s=0}^{n}(-1)^s\binom ns
 \binom{2Q+n}{Q+n-m-s}
 \bar u^{\,s}\bar v^{\,n-s}
 u^{Q+m+s}v^{Q+n-m-s},
 \label{eq:supp-harmonics}\\
 \mathcal N_{Qnm}
 ={}&\left[
 \frac{2Q+2n+1}{4\pi}
 \frac{(Q+n-m)!(Q+n+m)!}{n!(2Q+n)!}
 \right]^{1/2}.
 \label{eq:supp-harmonic-normalization}
\end{align}
Here $m=-Q-n,\ldots,Q+n$. A binomial coefficient is zero
when its lower argument lies outside its usual range; such terms are
omitted before evaluating spinor powers. The harmonics are orthonormal
with respect to $d\Omega=\sin\theta\,d\theta\,d\phi$.
In particular, the physical LLL orbitals are
\begin{equation}
 Y_{q,0,m}(z)=(-1)^{q-m}
 \sqrt{\frac{2q+1}{4\pi}\binom{2q}{q-m}}
 u(z)^{q+m}v(z)^{q-m}.
 \label{eq:supp-physical-lll}
\end{equation}
Their normalization follows from
\begin{equation}
 \int d\Omega\,|u|^{2a}|v|^{2b}
 =\frac{4\pi\,a!\,b!}{(a+b+1)!},
 \qquad a,b\in\mathbb N_0,
 \label{eq:supp-spinor-integral}
\end{equation}
and azimuthal orthogonality.

Let $\mathcal P_\kappa$ be the orthogonal projection onto the LLL
at monopole strength $\kappa$. If $h(u,v)$ is homogeneous of
degree $2\kappa+n$, then
\begin{equation}
 \mathcal P_\kappa\!\left[\bar u^{\,s}\bar v^{\,n-s}h(u,v)\right]
 =\frac{(2\kappa+1)!}{(2\kappa+n+1)!}
 \partial_u^s\partial_v^{n-s}h(u,v).
 \label{eq:supp-projection-rule}
\end{equation}
To verify this rule, take $h=u^av^b$ with $a+b=2\kappa+n$.
For $a\ge s$ and $b\ge n-s$, only the LLL monomial
$u^{a-s}v^{b-n+s}$ has the required azimuthal dependence.
Dividing its overlap with the unprojected monomial by its squared norm
using Eq.~\eqref{eq:supp-spinor-integral} gives
\begin{equation}
 \frac{(2\kappa+1)!}{(2\kappa+n+1)!}
 \frac{a!\,b!}{(a-s)!\,(b-n+s)!},
\end{equation}
which is the derivative coefficient in
Eq.~\eqref{eq:supp-projection-rule}. If either derivative exceeds
the corresponding monomial power, both sides vanish.

Applying this rule to $Y_{Q,n,m}J_i$ initially produces derivatives
acting on both $J_i$ and the explicit holomorphic powers in
Eq.~\eqref{eq:supp-harmonics}. The latter contributions cancel in the
sum over $s$. Indeed, fix $a$ derivatives with respect to $u$ and
$b$ with respect to $v$ to act on $J_i$, and set
$R=n-a-b$. For nonnegative remaining spinor powers, the factorials
in the Leibniz expansion obey
\begin{align}
 &\binom ns\binom sa\binom{n-s}b
 \binom{2Q+n}{Q+n-m-s}
 \frac{(Q+m+s)!(Q+n-m-s)!}{(Q+m+a)!(Q-m+b)!}
 \nonumber\\
 &\qquad=
 \frac{n!(2Q+n)!}{a!\,b!\,R!\,(Q+m+a)!(Q-m+b)!}
 \binom R{s-a}.
 \label{eq:supp-leibniz-factorials}
\end{align}
The factor outside the last binomial coefficient is independent of
$s$. The remaining alternating sum is
\begin{equation}
 \sum_{s=a}^{n-b}(-1)^s\binom{n-a-b}{s-a}
 =(-1)^a\sum_{j=0}^{R}(-1)^j\binom Rj
 =(-1)^a\delta_{R,0}.
 \label{eq:supp-leibniz-cancellation}
\end{equation}
Thus all $n$ derivatives can be placed to the right, where they act
only on $J_i$. If an implied remaining spinor power is negative,
the entire fixed-$(a,b)$ contribution vanishes before this
simplification.

\subsection{The shell convention used by the pairing amplitudes}
\label{app:JK}

The explicit coefficients in main-text Eq.~\eqref{eq:exact-p-coefficients} use the following
normalization of the JK projected orbitals \cite{KamillaJain,Kamilla_thesis,JainKamilla_IJMP}:
\begin{align}
 \widetilde Y_{nm;i}(Z)
 ={}&\frac{\mathcal N_{nm}}{J_i(Z)}
 \frac{(2N_1)!}{(2N_1+n)!}(-1)^{n+1/2-m}
 \sum_{s=0}^{n}(-1)^s\binom ns
 \binom{n+1}{n+1/2-m-s}
 \nonumber\\
 &\qquad\times
 u_i^{m+1/2+s}v_i^{n+1/2-m-s}
 \partial_{u_i}^{s}\partial_{v_i}^{n-s}J_i(Z),
 \label{eq:supp-jk-orbitals}\\
 \mathcal N_{nm}
 \equiv\mathcal N_{1/2,n,m}
 ={}&\sqrt{\frac{(n+\frac12-m)!(n+\frac12+m)!}
 {2\pi(n!)^2}}.
 \label{eq:supp-cf-normalization}
\end{align}
Here $u_i=u(z_i)$ and $v_i=v(z_i)$.

For clarity, this convention includes a shell-dependent rescaling
of the literal one-coordinate orthogonal projection. The product
$Y_{1/2,n,m}J_i$ has holomorphic degree $N_1+n$ and antiholomorphic
degree $n$ in its distinguished spinor, so its monopole strength is
$\kappa=N_1/2$. Eq.~\eqref{eq:supp-projection-rule} therefore
has the prefactor $(N_1+1)!/(N_1+n+1)!$ for that product.
The precise relation to Eq.~\eqref{eq:supp-jk-orbitals} is
\begin{equation}
 \widetilde Y_{nm;i}(Z)
 =\frac{(2N_1)!\,(N_1+n+1)!}
 {(2N_1+n)!\,(N_1+1)!}\,
 \frac{\mathcal P_{N_1/2}^{(i)}
 [Y_{1/2,n,m}(z_i)J_i(Z)]}{J_i(Z)}.
 \label{eq:supp-jk-projection-convention}
\end{equation}
Keeping this shell convention is necessary when using the numerical
values of $g_{p,n}$ stated in the main text. In general, a nonzero
shell rescaling $\widetilde Y_{nm;i}\mapsto a_n\widetilde Y_{nm;i}$
is absorbed by
\begin{equation}
 g_{p,n}\mapsto g_{p,n}/a_n^2,
 \qquad
 g_{s,n}\mapsto g_{s,n}/|a_n|^2.
 \label{eq:supp-shell-rescaling}
\end{equation}
It therefore leaves the variational families unchanged.

It is useful to introduce the polynomial numerators
\begin{equation}
 \mathcal F_{n,m;i}(Z)=J_i(Z)\widetilde Y_{nm;i}(Z).
 \label{eq:supp-numerators}
\end{equation}
Every term in Eq.~\eqref{eq:supp-jk-orbitals} differentiates $J_i$
exactly $n$ times. Since $J_i$ has degree $N_1-1$ in its distinguished
spinor, all projected orbitals with $n\ge N_1$ vanish.
For $0\le n\le N_1-1$, $\mathcal F_{n,m;i}$ is homogeneous of
degree $N_1$ in $z_i$ and degree one in every other spinor in $Z$.
These observations also make the polynomiality and the physical LLL
degree of the resulting many-body states explicit.

\section{\texorpdfstring{Details of the $s$-wave identity}{Details of the s-wave identity}}
\label{sec:supp-s}

For $g_{s,n}=\delta_{n,N_1-1}$, label the magnetic orbitals by
$r=q-m=0,\ldots,N_\phi$ and define
\begin{equation}
 F_{ir}(Z)=\mathcal F_{N_1-1,q-r;i}(Z),
 \qquad V_{ir}(Z)=Y_{q,0,q-r}(z_i).
 \label{eq:supp-s-matrices}
\end{equation}
All column subsets in this section are ordered by increasing $r$.
In particular,
\begin{equation}
 V_{ir}(Z)=(-1)^r\sqrt{\frac{N_1}{2\pi}\binom{N_\phi}{r}}
 u_i^{N_\phi-r}v_i^r.
 \label{eq:supp-s-V}
\end{equation}

\subsection{Generating polynomial for the highest shell}
\label{app:generating}
Introduce
\begin{equation}
 P_Z(t)=\prod_{k=1}^{N_1}(u_k+t v_k),
 \qquad E_s^{(i)}=[t^s]\prod_{k\ne i}(u_k+t v_k),
 \label{eq:supp-s-generators}
\end{equation}
where the notation $[t^s]$ means extracting the coefficient of $t^s$.
The Jastrow factor for the $i$th particle is $J_i(Z)=\prod_{j\ne i}(u_iv_j-u_jv_i)$. Replacing the distinguished spinor $(u_i,v_i)$ by $(u,v)$ gives
\begin{align}
 J_i(Z;u,v)
 &=\sum_{s=0}^{N_1-1}(-1)^{N_1-1-s}
 E_s^{(i)}u^s v^{N_1-1-s},
 \label{eq:supp-s-J-expansion}\\
 \partial_u^s\partial_v^{N_1-1-s}J_i(Z;u,v)
 &=(-1)^{N_1-1-s}s!(N_1-1-s)!E_s^{(i)}.
 \label{eq:supp-s-exhausted-derivatives}
\end{align}
The derivatives in the highest shell therefore remove all dependence
on the distinguished spinor from $J_i$.
Substituting $n=N_1-1$, $m=q-r$, and
Eq.~\eqref{eq:supp-s-exhausted-derivatives} into
Eq.~\eqref{eq:supp-jk-orbitals} gives
\begin{align}
 F_{ir}(Z)
 ={}&(-1)^{N_1-1+r}\frac{(2N_1)!}{(3N_1-1)!}
 \sqrt{\frac{r!(N_\phi-r)!}{2\pi}}
 \sum_{s=0}^{N_1-1}\binom{N_1}{r-s}E_s^{(i)}
 u_i^{N_1-r+s}v_i^{r-s}.
 \label{eq:supp-s-expanded-F}
\end{align}
The factors $s!(N_1-1-s)!$ cancel the denominator of
$\binom{N_1-1}{s}$, and the two $s$-dependent signs give
$(-1)^{N_1-1}$.
The remaining sum is
\begin{align}
 \sum_{s=0}^{N_1-1}\binom{N_1}{r-s}E_s^{(i)}
 u_i^{N_1-r+s}v_i^{r-s}
 &=[t^r]\left[(u_i+t v_i)^{N_1}
 \prod_{k\ne i}(u_k+t v_k)\right]
 \nonumber\\
 &=[t^r]\left[(u_i+t v_i)^{N_1-1}P_Z(t)\right].
\end{align}
Thus the exact form of main-text Eq.~\eqref{eq:s-generating-orbitals} is
\begin{align}
 F_{ir}(Z)
 ={}&(-1)^{N_1-1}\frac{(2N_1)!}{(3N_1-1)!}
 \sqrt{\frac{N_\phi!}{2\pi}}
 \frac{(-1)^r}{\sqrt{\binom{N_\phi}{r}}}
 [t^r]\left[(u_i+t v_i)^{N_1-1}P_Z(t)\right].
 \label{eq:supp-s-generating}
\end{align}
The Fourier representation in the main text follows by setting $t=e^{-i\phi}$: its spinor product is $e^{iq\phi}(u_i+e^{-i\phi}v_i)^{N_1-1}P_Z(e^{-i\phi})$, so multiplication by $e^{-im\phi}$ with $m=q-r$ makes the integral extract the coefficient of $t^r$.

\subsection{Bilinear orthogonality and complementary minors}
\label{app:orthogonality}
Write
\begin{equation}
 a_{ir}=[t^r]\left[(u_i+t v_i)^{N_1-1}P_Z(t)\right],
 \qquad C_{rs}=(-1)^r\delta_{r+s,N_\phi}.
\end{equation}
The binomial weights in Eqs.~\eqref{eq:supp-s-V} and
\eqref{eq:supp-s-generating} cancel in $FCV^T$. Apart from one
common nonzero scalar, its entries are
\begin{align}
 (FCV^T)_{ij}
 &\propto\sum_{r=0}^{N_\phi}(-1)^r a_{ir}u_j^rv_j^{N_\phi-r}
 \nonumber\\
 &=(u_i v_j-v_i u_j)^{N_1-1}
 \prod_{k=1}^{N_1}(u_kv_j-v_ku_j)=0.
 \label{eq:supp-s-bilinear}
\end{align}
The final product contains the vanishing factor $k=j$.
This polynomial argument covers the poles without dividing by a
spinor component. In particular, the transpose here is an ordinary
transpose, with no complex conjugation.

For an $N_1$-element subset $R$ of $\{0,\ldots,N_\phi\}$,
let $R^c$ be its complement and
$\rho(R^c)=\{N_\phi-r:r\in R^c\}$, sorted increasingly.
Set $W=VC^T$, so that $FW^T=0$ and
\begin{equation}
 W_{ir}=(-1)^rV_{i,N_\phi-r}.
 \label{eq:supp-s-W}
\end{equation}
For generic coordinates $F$ and $W$ both have rank $N_1$.
Their bilinearly orthogonal row spaces imply
\begin{equation}
 \det F_R=\lambda(Z)\epsilon(R)\det W_{R^c},
 \qquad
 \epsilon(R)=(-1)^{\sum_{r\in R}r-N_1(N_1-1)/2},
 \label{eq:supp-s-complementary-minors}
\end{equation}
where $\lambda(Z)$ is independent of $R$.

To see the matrix identity, work on a patch where the first $N_1$
columns of $F$ form an invertible matrix. Invertible row operations
put $F$ into the form $(I,T)$. Orthogonality and full rank then put
$W$ into the form $(-T^T,I)$, up to a second invertible row operation.
Applying the Jacobi's complementary-minor
identity~\cite{BrualdiSchneider1983} to the unit-determinant
block matrix and its inverse,
\begin{equation}
 B=\begin{pmatrix}I&T\\0&I\end{pmatrix},
 \qquad B^{-1}=\begin{pmatrix}I&-T\\0&I\end{pmatrix},
\end{equation}
gives
\begin{equation}
 \det(I,T)_R=\epsilon(R)\det(-T^T,I)_{R^c}.
\end{equation}
The two row transformations supply the single factor $\lambda(Z)$.
The sign can also be read directly from a Laplace expansion:
if $R=\{r_1<\cdots<r_{N_1}\}$, moving the selected columns to
the first $N_1$ positions requires
\begin{equation}
 \sum_{a=1}^{N_1}[r_a-(a-1)]
 =\sum_{r\in R}r-\frac{N_1(N_1-1)}2
 \label{eq:supp-shuffle-transpositions}
\end{equation}
transpositions. This proves the sign in
Eq.~\eqref{eq:supp-s-complementary-minors}.

Eq.~\eqref{eq:supp-s-W} supplies
$(-1)^{\sum_{r\in R^c}r}$. Reflecting the selected indices reverses
their order, so sorting them requires another
$N_1(N_1-1)/2$ transpositions. Consequently,
\begin{equation}
 \det W_{R^c}
 =(-1)^{\sum_{r\in R^c}r+N_1(N_1-1)/2}
 \det V_{\rho(R^c)}.
\end{equation}
Multiplying by $\epsilon(R)$ leaves
$(-1)^{\sum_{r=0}^{N_\phi}r}=(-1)^{N_1}$, independent of $R$.
Absorbing that common sign into $\lambda(Z)$ yields
\begin{equation}
 \det F_R(Z)=\beta(Z)\det V_{\rho(R^c)}(Z).
 \label{eq:supp-s-minor-ratio}
\end{equation}

\subsection{Evaluation of the common factor}
\label{app:evaluation}
Take $R_0=\{0,\ldots,N_1-1\}$, for which
$\rho(R_0^c)=R_0$, and first set $u_i=1$.
Write $P_Z(t)=\sum_{j=0}^{N_1}p_jt^j$, with $p_0=1$.
For $r\in R_0$,
\begin{equation}
 a_{ir}=\sum_{k=0}^{r}\binom{N_1-1}{k}v_i^k p_{r-k}.
\end{equation}
Thus $(a_{ir})_{r\in R_0}=LT$, where
$L_{ik}=\binom{N_1-1}{k}v_i^k$ and
$T_{kr}=p_{r-k}$ for $r\ge k$, with zero entries otherwise.
The matrix $T$ is upper triangular with unit diagonal. Hence
\begin{equation}
 \det(a_{ir})_{r\in R_0}
 =\left[\prod_{r=0}^{N_1-1}\binom{N_1-1}{r}\right]
 \prod_{i<j}(v_j-v_i).
 \label{eq:supp-s-vandermonde}
\end{equation}
This also proves the generic rank statement used above.
The corresponding minor of $V$ contains the same Vandermonde
determinant. Restoring the column weights and row factors gives
\begin{equation}
 \boxed{\displaystyle
 \beta(Z)=\beta_{N_1}
 =\left[\frac{(2N_1)!}{(3N_1-1)!}
 \sqrt{\frac{(2N_1-1)!}{N_1}}\right]^{N_1}
 \prod_{r=0}^{N_1-1}
 \frac{\binom{N_1-1}{r}}{\binom{2N_1-1}{r}}>0.}
 \label{eq:supp-beta}
\end{equation}
The $(-1)^r$ column signs cancel in the ratio, and the row sign
$(-1)^{N_1-1}$ in $F$ contributes
$(-1)^{N_1(N_1-1)}=1$ to its determinant.

Every maximal minor of $F$ has degree $N_\phi$ in each particle
spinor: a given spinor has degree $N_1$ in its own row and degree
one in each of the other $N_1-1$ rows. The same homogeneity holds
for the minors of $V$. Rescaling each spinor therefore extends the
constant ratio from $u_i=1$ to the patch $u_i\ne0$.
Polynomiality extends it to all remaining configurations.
We have established the exact version of main-text Eq.~\eqref{eq:s-complementary-minors}:
\begin{equation}
 \boxed{\det F_R(Z)=\beta_{N_1}\det V_{\rho(R^c)}(Z).}
 \label{eq:supp-s-exact-minors}
\end{equation}

\subsection{The many-body identity}

For either layer,
\begin{equation}
 \prod_{i=1}^{N_1}J_i(Z)
 =(-1)^{N_1(N_1-1)/2}J(Z)^2,
 \label{eq:supp-J-sign}
\end{equation}
because each unordered pair contributes
$[z_i,z_j][z_j,z_i]=-[z_i,z_j]^2$.
The signs from the electron and hole layers cancel, so
\begin{equation}
 \Phi_{\mathrm{BCS},s}^{\mathrm{ph}}(\boldsymbol g_s;X,H)
 =\det[F(X)F(H)^\dagger].
\end{equation}
Cauchy--Binet~\cite{BrualdiSchneider1983} and Eq.~\eqref{eq:supp-s-exact-minors} give
\begin{align}
 \Phi_{\mathrm{BCS},s}^{\mathrm{ph}}(\boldsymbol g_s;X,H)
 &=\sum_{|R|=N_1}\det F_R(X)\det F_R(H)^*
 \nonumber\\
 &=\beta_{N_1}^{\,2}\sum_{|R|=N_1}
 \det V_{\rho(R^c)}(X)\det V_{\rho(R^c)}(H)^*
 \nonumber\\
 &=\beta_{N_1}^{\,2}\det[V(X)V(H)^\dagger]
 =\beta_{N_1}^{\,2}\Phi_{111}^{\mathrm{ph}}(X,H).
 \label{eq:supp-s-identity}
\end{align}
The third equality uses the fact that $R\mapsto\rho(R^c)$ is a
bijection on the $N_1$-element subsets. Complex conjugation appears
here in the hole-layer determinant; it was not used in the bilinear
minor relation.

\section{\texorpdfstring{Details of the $p$-wave identity}{Details of the p-wave identity}}
\label{sec:supp-p}

\subsection{The polynomial kernel at two distinguished poles}
\label{app:distinguished_poles}
Using Eq.~\eqref{eq:supp-numerators}, the kernel of main-text
Eq.~\eqref{eq:p-polynomial-kernel} is
\begin{equation}
 K_{ij}=\sum_{n=0}^{N_1-1}g_{p,n}
 \sum_{m=-n-1/2}^{n+1/2}(-1)^{m+1/2}
 \mathcal F_{n,m;i}(X)\mathcal F_{n,-m;j}(Y).
 \label{eq:supp-p-K}
\end{equation}
First set the distinguished spinors to
$(u(x_i),v(x_i))=(1,0)$ and
$(u(y_j),v(y_j))=(0,1)$, while keeping all spectators arbitrary.
Define their coefficients by
\begin{align}
 J_i(X;u,v)&=\sum_{r=0}^{N_1-1}(-1)^r A_r u^{N_1-1-r}v^r,
 \label{eq:supp-p-A}\\
 J_j(Y;u,v)&=\sum_{r=0}^{N_1-1}(-1)^{N_1-1-r}B_r u^r v^{N_1-1-r}.
 \label{eq:supp-p-B}
\end{align}
In particular,
\begin{equation}
 A_{N_1-1}=\prod_{k\ne i}u(x_k),
 \qquad B_{N_1-1}=\prod_{\ell\ne j}v(y_\ell).
 \label{eq:supp-p-last-coefficients}
\end{equation}

At the north pole, a surviving term in
Eq.~\eqref{eq:supp-jk-orbitals} must have zero explicit power of
$v_i$. This requires $m=r+1/2$ and $s=n-r$, with $0\le r\le n$.
The derivative is
\begin{equation}
 \left.\partial_u^{n-r}\partial_v^rJ_i(X;u,v)\right|_{(1,0)}
 =(-1)^r r!\frac{(N_1-1-r)!}{(N_1-1-n)!}A_r.
\end{equation}
Combining it with the binomial coefficient and
$\mathcal N_{n,r+1/2}$ gives
\begin{equation}
 \left.\mathcal F_{n,r+1/2;i}(X)\right|_{\mathrm{poles}}
 =(-1)^r\frac{(2N_1)!}{(2N_1+n)!}
 \frac{(N_1-1-r)!}{(N_1-1-n)!}
 \sqrt{\frac{(n+r+1)!}{2\pi(n-r)!}}A_r.
 \label{eq:supp-p-north}
\end{equation}
For the paired lower-layer orbital, $m=-r-1/2$. At the south pole,
zero explicit power of $u$ requires $s=r$. Its derivative is
\begin{equation}
 \left.\partial_u^r\partial_v^{n-r}J_j(Y;u,v)\right|_{(0,1)}
 =(-1)^{N_1-1-r}r!\frac{(N_1-1-r)!}{(N_1-1-n)!}B_r,
\end{equation}
and the full numerator is
\begin{equation}
 \left.\mathcal F_{n,-r-1/2;j}(Y)\right|_{\mathrm{poles}}
 =(-1)^{n+N_1-r}\frac{(2N_1)!}{(2N_1+n)!}
 \frac{(N_1-1-r)!}{(N_1-1-n)!}
 \sqrt{\frac{(n+r+1)!}{2\pi(n-r)!}}B_r.
 \label{eq:supp-p-south}
\end{equation}
The two orbital signs and the pairing sign $(-1)^{r+1}$ multiply
to $(-1)^{N_1-1+n+r}$. Substitution into the kernel yields
\begin{align}
 \left.K_{ij}\right|_{\mathrm{poles}}
 ={}&\frac{1}{2\pi}\sum_{r=0}^{N_1-1}
 (-1)^{N_1-1+r}(N_1-1-r)!^2 A_rB_r
 \nonumber\\
 &\times\sum_{n=r}^{N_1-1}(-1)^n g_{p,n}
 \left[\frac{(2N_1)!}{(2N_1+n)!}\right]^2
 \frac{(n+r+1)!}{(n-r)!(N_1-1-n)!^2}.
 \label{eq:supp-p-pole-kernel}
\end{align}

\subsection{The finite-sum cancellation}
\label{app:telescoping}
Insert the main-text coefficients
\begin{equation}
 g_{p,n}=(n+1)
 \frac{(N_1-1-n)!}{(N_1+n+1)!}
 \left[\frac{(2N_1+n)!}{(2N_1)!}\right]^2.
 \label{eq:supp-p-coefficients}
\end{equation}
The squared factorial ratio cancels the factors from the two
projected orbitals, leaving
\begin{align}
 \left.K_{ij}\right|_{\mathrm{poles}}
 &=\frac{1}{2\pi}
 \sum_{r=0}^{N_1-1}(-1)^{N_1-1+r}
 (N_1-1-r)!^2 A_rB_r S_r,
 \label{eq:supp-p-pole-S}\\
 S_r&=\sum_{n=r}^{N_1-1}(-1)^n
 \frac{(n+1)(n+r+1)!}
 {(n-r)!(N_1-1-n)!(N_1+n+1)!}.
 \label{eq:supp-p-S}
\end{align}
For $r<N_1-1$, introduce
\begin{equation}
 T_n=(-1)^n\frac{(n+r+2)!}
 {(n-r)!(N_1-2-n)!(N_1+n+1)!},
 \qquad r\le n\le N_1-2,
\end{equation}
with $T_{r-1}=T_{N_1-1}=0$. Direct subtraction gives
\begin{equation}
 T_n-T_{n-1}=2(N_1-1-r)(-1)^n
 \frac{(n+1)(n+r+1)!}
 {(n-r)!(N_1-1-n)!(N_1+n+1)!}.
 \label{eq:supp-p-telescoping}
\end{equation}
After extracting the common factorial ratio, the required numerator
identity is simply
\begin{equation}
 (n+r+2)(N_1-1-n)+(n-r)(N_1+n+1)
 =2(N_1-1-r)(n+1).
\end{equation}
The boundary prescription makes Eq.~\eqref{eq:supp-p-telescoping}
valid at both endpoints. Therefore
\begin{equation}
 S_r=\frac{T_{N_1-1}-T_{r-1}}{2(N_1-1-r)}=0,
 \qquad r<N_1-1.
\end{equation}
For $r=N_1-1$, only one term remains:
\begin{equation}
 S_{N_1-1}=(-1)^{N_1-1}
 \frac{N_1(2N_1-1)!}{(2N_1)!}
 =\frac{(-1)^{N_1-1}}2.
 \label{eq:supp-p-S-last}
\end{equation}
This branch also covers $N_1=1$.
Only the last spectator coefficients survive in
Eq.~\eqref{eq:supp-p-pole-S}, giving
\begin{equation}
 \left.K_{ij}\right|_{\mathrm{poles}}
 =(-1)^{N_1-1}\frac{1}{4\pi}
 \prod_{k\ne i}u(x_k)\prod_{\ell\ne j}v(y_\ell).
 \label{eq:supp-p-poles-final}
\end{equation}

\subsection{Extension from the poles to arbitrary spinors}
\label{app:extension}
We now prove the full polynomial identity
\begin{equation}
 K_{ij}=c_{N_1}[x_i,y_j]
 \prod_{k\ne i}[x_k,y_j]\prod_{\ell\ne j}[x_i,y_\ell],
 \qquad c_{N_1}=(-1)^{N_1-1}\frac{1}{4\pi}.
 \label{eq:supp-p-factorization}
\end{equation}
Both sides have degree $N_1$ in each distinguished spinor and
degree one in every spectator spinor. They are invariant under
simultaneous $SU(2)$ rotations. For the kernel, this follows from
the covariance of the projected orbitals and the singlet contraction
of each shell; the pairing phase differs from the usual singlet
phase only by a shell-dependent sign,
$(-1)^{m+1/2}=(-1)^{n+1}(-1)^{\ell_n-m}$.

For holomorphic polynomials, $SU(2)$ invariance extends to
$SL(2,\mathbb C)$: annihilation by the $\mathfrak{su}(2)$ generators
implies annihilation by their complex linear combinations.
Let
\begin{equation}
 S=\begin{pmatrix}u(x_i)&u(y_j)\\v(x_i)&v(y_j)\end{pmatrix},
 \qquad w=\det S=[x_i,y_j].
\end{equation}
For $w\ne0$, define
\begin{equation}
 L=\begin{pmatrix}1&0\\0&w\end{pmatrix}S^{-1},
 \qquad \det L=1.
 \label{eq:supp-p-SL2-map}
\end{equation}
We apply this same matrix simultaneously to every particle spinor in both layers, including all spectators:
\begin{equation}
 \begin{pmatrix}u(z)\\v(z)\end{pmatrix}
 \longmapsto
 L\begin{pmatrix}u(z)\\v(z)\end{pmatrix},
 \qquad z\in X\cup Y.
\end{equation}
Both sides of the proposed polynomial identity are unchanged by this common $SL(2,\mathbb C)$ transformation. Since the columns of $S$ are the two distinguished spinors, the identity $LS=\operatorname{diag}(1,w)$ shows that their images are $(1,0)$ and $(0,w)$. The spectator spinors are transformed as well and remain arbitrary. Homogeneity in the second distinguished spinor then reduces the comparison to $(1,0)$ and $(0,1)$, with a common factor $w^{N_1}$ on both sides. The remaining comparison is
exactly Eq.~\eqref{eq:supp-p-poles-final}; the transformed spectator
spinors remain arbitrary. Polynomiality extends the identity from
$w\ne0$ to $w=0$. A physical rotation alone would not suffice to
map a generic pair of points to opposite poles; the complexified
transformation is what makes this reduction valid.

\subsection{Cauchy determinant and the overall phase}
\label{app:Cauchy}
Eq.~\eqref{eq:supp-J-sign} shows that the two layer signs
cancel and $\Phi_{\mathrm{BCS},p}(\boldsymbol g_p;X,Y)=\det K$.
Define
\begin{equation}
 P_i=\prod_{\ell=1}^{N_1}[x_i,y_\ell],
 \qquad Q_j=\prod_{k=1}^{N_1}[x_k,y_j].
\end{equation}
The kernel factorization reads
$K_{ij}=c_{N_1}P_i[x_i,y_j]^{-1}Q_j$.
Factoring $P_i$ from row $i$ and $Q_j$ from column $j$ gives
\begin{equation}
 \det K=c_{N_1}^{N_1}
 \left(\prod_{i,j=1}^{N_1}[x_i,y_j]\right)^2
 \det_{i,j}\left[\frac{1}{[x_i,y_j]}\right].
\end{equation}
With the bracket convention used here, the spherical Cauchy identity is
\begin{equation}
 \det_{i,j}\left[\frac{1}{[x_i,y_j]}\right]
 =(-1)^{N_1(N_1-1)/2}
 \frac{J(X)J(Y)}{\prod_{i,j=1}^{N_1}[x_i,y_j]}.
 \label{eq:supp-cauchy}
\end{equation}
The sign can be fixed in the affine patch $x_i=(1,z_i)$,
$y_j=(1,w_j)$, where $[x_i,y_j]=w_j-z_i$.
Multiplying the determinant by its common denominator gives a
polynomial antisymmetric separately in the $z_i$ and $w_j$.
Its degree is saturated by the two Vandermonde products, so the
remaining factor is a constant $C_{N_1}$.
Taking $z_{N_1}\to w_{N_1}$ and evaluating the surviving cofactor
gives $C_{N_1}=(-1)^{N_1-1}C_{N_1-1}$, with $C_1=1$.
Thus $C_{N_1}=(-1)^{N_1(N_1-1)/2}$, and homogeneity restores
Eq.~\eqref{eq:supp-cauchy} for general spinors.

Since the sign in $c_{N_1}^{N_1}$ is
$(-1)^{N_1(N_1-1)}=1$, the exact wavefunction identity is
\begin{equation}
 \boxed{\displaystyle
 \Phi_{\mathrm{BCS},p}(\boldsymbol g_p;X,Y)
 =(-1)^{N_1(N_1-1)/2}
 \left[\frac{1}{4\pi}\right]^{N_1}
 \Phi_{111}(X,Y).}
 \label{eq:supp-p-identity}
\end{equation}
The apparent denominators cancel, so the final identity also holds
when interlayer coordinates coincide.

\section{Particle--hole transformation and normalization}
\label{sec:supp-normalization}

\subsection{Filled LLL determinant}

We take coordinate inner products with the product angular measure,
without factorial prefactors:
\begin{equation}
 \|\Phi\|^2=\int\prod_{i=1}^{N_1}d\Omega_{x_i}
 \prod_{j=1}^{N_1}d\Omega_{y_j}\,|\Phi(X,Y)|^2,
 \label{eq:supp-measure}
\end{equation}
and analogously in electron--hole coordinates.
For $D$ orbitals at flux $D-1$, define
\begin{equation}
 A_D=\left(\frac{D}{4\pi}\right)^{D/2}
 \sqrt{\prod_{r=0}^{D-1}\binom{D-1}{r}}.
 \label{eq:supp-A}
\end{equation}
Order the physical LLL orbitals by increasing $m$, and write
$\varphi_a(z)=Y_{q,0,-q+a}(z)$ for $a=0,\ldots,2N_1-1$.
For $Z_e=(x_1,\ldots,x_{N_1},y_1,\ldots,y_{N_1})$,
\begin{equation}
 \mathcal D(X,Y)
 =\det_{\alpha,a}[\varphi_a((Z_e)_\alpha)]
 =A_{2N_1}\Phi_{111}(X,Y).
 \label{eq:supp-filled-determinant}
\end{equation}
To check the sign, first order monomials by increasing
$r=q-m$. Their phases contribute
$(-1)^{(2N_1)(2N_1-1)/2}$, and reversing the column order to
increasing $m$ contributes the same sign. They cancel, leaving
the positive bracket Vandermonde in Eq.~\eqref{eq:supp-filled-determinant}.
The $V$ matrices in App.~\ref{sec:supp-s} instead use increasing
$r$; reversing all orbital columns leaves their paired kernels unchanged.

A determinant of $2N_1$ orthonormal orbitals has squared norm
$(2N_1)!$. Hence
\begin{equation}
 \|\Phi_{111}\|^2=\frac{(2N_1)!}{A_{2N_1}^{\,2}},
 \qquad
 \Psi_{111}=\frac{A_{2N_1}}{\sqrt{(2N_1)!}}\Phi_{111}.
 \label{eq:supp-111-normalized}
\end{equation}

\subsection{An explicit change to hole coordinates}

Fix the overall particle--hole phase by defining
\begin{equation}
 (\mathcal P_\downarrow\Phi)(X,H)
 =\frac{1}{N_1!}\int\prod_{j=1}^{N_1}d\Omega_{y_j}\,
 \mathcal D(H,Y)^*\Phi(X,Y),
 \label{eq:supp-ph-map}
\end{equation}
where the holes are placed first in the coordinate order of
$\mathcal D(H,Y)$. This is a linear change from the lower-layer
electron occupation basis to the complementary hole occupation
basis. Complex conjugation in the kernel supplies the conjugate
hole orbitals; it does not act on the coefficients of $\Phi$.

Let $\mathcal V_R(Z)$ be the determinant of the orbitals
$\varphi_a$ in an $N_1$-element set $R$, ordered increasingly.
The Laplace expansion along the first $N_1$ coordinate rows is
\begin{equation}
 \mathcal D(X,Y)=\sum_{|R|=N_1}
 \epsilon(R)\mathcal V_R(X)\mathcal V_{R^c}(Y),
 \qquad
 \epsilon(R)=(-1)^{\sum_{r\in R}r-N_1(N_1-1)/2}.
 \label{eq:supp-filled-laplace}
\end{equation}
Here the integer labels refer to increasing $m$.
The sign is the shuffle parity calculated in
Eq.~\eqref{eq:supp-shuffle-transpositions}.
Orbital orthogonality implies
\begin{equation}
 \int\prod_{j=1}^{N_1}d\Omega_{y_j}\,
 \mathcal V_R(Y)^*\mathcal V_S(Y)=N_1!\,\delta_{R,S}.
 \label{eq:supp-minor-orthogonality}
\end{equation}
Inserting the two Laplace expansions into
Eq.~\eqref{eq:supp-ph-map}, orthogonality sets their subsets
equal and the two factors $\epsilon(R)$ multiply to one. Thus
\begin{align}
 \mathcal P_\downarrow\mathcal D(X,Y)
 &=\sum_{|R|=N_1}\mathcal V_R(X)\mathcal V_R(H)^*
 \nonumber\\
 &=\det_{i,j}[K_q(x_i,h_j)]
 =\Phi_{111}^{\mathrm{ph}}(X,H),
 \label{eq:supp-ph-filled}
\end{align}
where Cauchy--Binet was used and
\begin{align}
 K_q(x,h)
 &=\sum_{m=-q}^{q}Y_{q,0,m}(x)Y_{q,0,m}(h)^*
 \nonumber\\
 &=\frac{2q+1}{4\pi}
 [u(x)u(h)^*+v(x)v(h)^*]^{2q}.
 \label{eq:supp-lll-kernel}
\end{align}
This proves main-text Eq.~\eqref{eq:111-ph}, including the coordinate-independent
factor suppressed there:
\begin{equation}
 \mathcal P_\downarrow\Phi_{111}
 =A_{2N_1}^{-1}\Phi_{111}^{\mathrm{ph}}.
 \label{eq:supp-ph-polynomial}
\end{equation}

More generally, the same map acts on a normalized lower-layer Slater
determinant as
\begin{equation}
 \mathcal P_\downarrow\frac{\mathcal V_S(Y)}{\sqrt{N_1!}}
 =\epsilon(S^c)\frac{\mathcal V_{S^c}(H)^*}{\sqrt{N_1!}}.
 \label{eq:supp-ph-basis}
\end{equation}
Complementation permutes an orthonormal basis, so the transformation
is invertible and norm preserving. This justifies reversing it in
the $s$-wave identity.

The sum in Eq.~\eqref{eq:supp-ph-filled} contains
$\binom{2N_1}{N_1}$ mutually orthogonal terms, each of squared
norm $(N_1!)^2$. Therefore
\begin{equation}
 \|\Phi_{111}^{\mathrm{ph}}\|^2=(2N_1)!,
 \qquad
 \Psi_{111}^{\mathrm{ph}}
 =\frac{\Phi_{111}^{\mathrm{ph}}}{\sqrt{(2N_1)!}},
 \qquad
 \mathcal P_\downarrow\Psi_{111}=\Psi_{111}^{\mathrm{ph}}.
 \label{eq:supp-111ph-normalized}
\end{equation}

\subsection{Normalization of the paired states}
\label{app:normalization}

For the coefficient choices in the main text, the exact constants
in Eqs.~\eqref{eq:supp-s-identity} and \eqref{eq:supp-p-identity}
give
\begin{align}
 \|\Phi_{\mathrm{BCS},p}(\boldsymbol g_p)\|^2
 &=\left[\frac{1}{4\pi}\right]^{2N_1}
 \frac{(2N_1)!}{A_{2N_1}^{\,2}},
 \label{eq:supp-p-norm}\\
 \|\Phi_{\mathrm{BCS},s}^{\mathrm{ph}}(\boldsymbol g_s)\|^2
 &=\beta_{N_1}^{\,4}(2N_1)!.
 \label{eq:supp-s-norm}
\end{align}
The positive number $\beta_{N_1}$ is defined in
Eq.~\eqref{eq:supp-beta}. Choosing the overall phase of the
$p$-wave state consistently, the normalized paired wavefunctions are
\begin{align}
 \Psi_{\mathrm{BCS},p}(\boldsymbol g_p;X,Y)
 &=\frac{(-1)^{N_1(N_1-1)/2}A_{2N_1}}{\sqrt{(2N_1)!}}
 \left[4\pi\right]^{N_1}
 \Phi_{\mathrm{BCS},p}(\boldsymbol g_p;X,Y)
 =\Psi_{111}(X,Y),
 \label{eq:supp-p-normalized}\\
 \Psi_{\mathrm{BCS},s}^{\mathrm{ph}}(\boldsymbol g_s;X,H)
 &=\frac{\Phi_{\mathrm{BCS},s}^{\mathrm{ph}}(\boldsymbol g_s;X,H)}
 {\beta_{N_1}^{\,2}\sqrt{(2N_1)!}}
 =\Psi_{111}^{\mathrm{ph}}(X,H).
 \label{eq:supp-s-normalized}
\end{align}
Equivalently, in electron coordinates,
\begin{equation}
 \mathcal P_\downarrow^{-1}
 \Psi_{\mathrm{BCS},s}^{\mathrm{ph}}(\boldsymbol g_s)
 =\Psi_{111}
 =\Psi_{\mathrm{BCS},p}(\boldsymbol g_p).
\end{equation}
If the coordinate inner product instead includes
$(N_1!)^{-2}$ multiplying the product measure, every unit-normalized
wavefunction above is multiplied by $N_1!$.
This common convention change does not affect either identity.
All formulas include $N_1=1$, with empty products interpreted as one.

\section{CF occupations for the $p$-wave state that reproduces the (111) state}
\label{app:p_wave_occupations}

In the main text we derive the expression for the wavefunction parameters $g_{p,n}$ that reproduce the Halperin (111) state from the $p$-wave paired CFs. However, to understand how CF shells $n$ contribute most to the wavefunction, it is helpful to compute the occupation probability $p_n$ of the $n$th shell. $p_n=1$ means that all $2(n+1)$ orbitals in the $n$th shell are filled. The expression for $p_n$ for the $p$-wave paired CF wavefunction was introduced in Ref.~\cite{Simon1} and was used for the $s$-wave paired CF wavefunction in Ref.~\cite{Hu2024single}. In this section we derive the analytical expression for $p_n$.

For brevity, write $N=N_1$ and
$\Phi(\mathbf g)=\Phi_{\mathrm{BCS},p}(\mathbf g)$.
Using the unnormalized projected state, we define the effective
occupation per magnetic orbital in shell $n$ by \cite{Simon1}
\begin{equation}
 p_n=\left.
 \frac{1}{2d_n}\frac{\partial\ln\mathcal Z(\mathbf g)}
 {\partial\ln g_n}\right|_{\mathbf g=\mathbf g_p},
 \qquad
 \mathcal Z(\mathbf g)=\langle\Phi(\mathbf g)|\Phi(\mathbf g)\rangle,
 \quad d_n=2n+2.
 \label{eq:occ_definition}
\end{equation}
The derivative is taken for real positive $g_n$, with the other
amplitudes fixed, before setting $\mathbf g=\mathbf g_p$.
This definition is invariant under constant rescalings of the CF
shells with compensating rescalings of $g_n$, but characterizes the
chosen projected variational representation.

Absorbing the individual Jastrow factors gives
$\Phi=\det K$, where
\begin{equation}
 K=\sum_{n=0}^{N-1}g_n\mathcal K^{(n)},
 \qquad
 \mathcal K^{(n)}_{ij}
 =\sum_{m=-n-1/2}^{n+1/2}(-1)^{m+1/2}
 F_{n,m;i}(X)F_{n,-m;j}(Y),
 \label{eq:occ_shell_kernel}
\end{equation}
and $F_{n,m;i}=J_i\widetilde Y_{nm;i}$.
The factorized kernel at $\mathbf g_p$ and the Cauchy inverse give
\begin{align}
 L_n(X,Y)
 &\equiv\left.g_n\partial_{g_n}\ln\Phi\right|_{\mathbf g_p}
 =g_{p,n}\operatorname{tr}\!\left[(K^\star)^{-1}\mathcal K^{(n)}\right]
 \notag\\
 &=4\pi g_{p,n}
 \sum_{i,j}
 \frac{\mathcal K^{(n)}_{ij}}
 {[x_i,y_j]J_i(X)J_j(Y)},
 \qquad K^\star=K(\mathbf g_p).
 \label{eq:occ_local_response}
\end{align}
Eq.~\eqref{eq:occ_definition} therefore reads
$d_np_n=\operatorname{Re}\langle L_n\rangle$, with the average
taken in the exact $(111)$ state.

\begin{figure}
    \centering
    \includegraphics[width=0.5\linewidth]{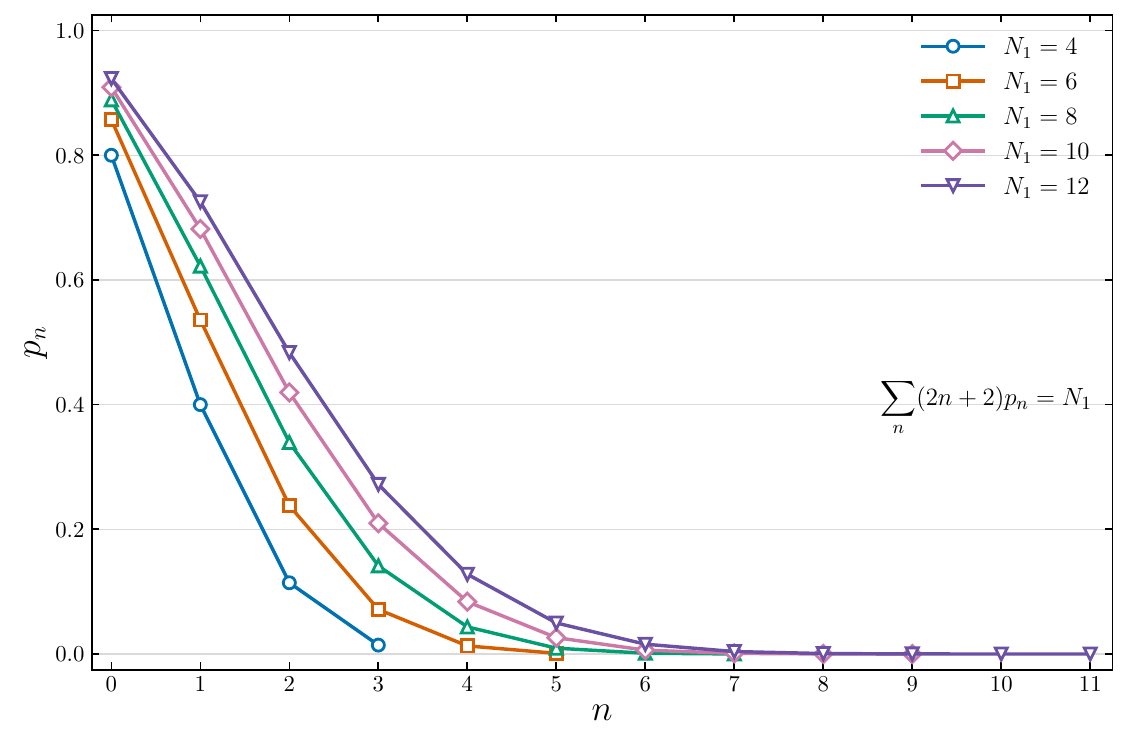}
    \caption{\textbf{Occupation probabilities of CF orbitals for $p$-wave trial state.} We show the occupation probabilities of the CF orbitals for the BCS wavefunction that exactly reproduces the Halperin (111) state. All CF orbitals have non-zero occupation probability, with the lower $n$ orbitals having larger occupation probabilities. This is the opposite of the $s$-wave wave function representation of the Halperin (111) state, where only the highest CF shell is occupied. }
    \label{fig:cf_occupations_vs_n}
\end{figure}

In fact, $L_n$ is coordinate independent. Each shell kernel
$\mathcal K^{(n)}_{ij}$ vanishes at $x_i=y_j$: rotating the
coincident points to the north pole, only $m>0$ projected orbitals
survive, whereas the contraction pairs $m$ with $-m$.
At this coincidence, $K^\star$ has a zero $i$th row and $j$th
column. Its cofactor expansion then implies
$\left.\partial_{g_n}\det K\right|_{\mathbf g_p}=0$ there.
This derivative is also antisymmetric within each layer and has
degree $2N-1$ in every spinor. It is consequently divisible by
$\Phi_{111}$, which already saturates these degrees, and hence
is proportional to $\Phi_{111}$. Thus $L_n$ can be evaluated
at any generic configuration.

Choose affine spinors $(1,x_i)$ and $(1,y_j)$, with $x_i$ and
$y_j$ the roots of $P_X(z)=z^N-a$ and $P_Y(z)=z^N-b$,
respectively, where $ab(a-b)\ne0$. The differentiated Jastrow
factor is
\begin{equation}
 \left.\partial_u^s\partial_v^{n-s}J_i(X;u,v)\right|_{(1,x_i)}
 =(-1)^{N-1}s!(n-s)!\binom{N}{n+1}
 x_i^{N-1-n+s}.
 \label{eq:occ_root_derivative}
\end{equation}
Substituting into the differential JK orbitals and combining the
magnetic labels $m$ and $-m$ yields
\begin{align}
 \mathcal K^{(n)}(x,y)
 &=\frac{H_n^2}{2\pi}\binom{N}{n+1}^{\!2}
 \sum_{r=0}^{n}c_{nr}
 \left(x^{\alpha_r}y^{\beta_r}-x^{\beta_r}y^{\alpha_r}\right),
 \label{eq:occ_root_kernel}\\
 H_n&=\frac{(2N)!}{(2N+n)!},\qquad
 c_{nr}=(-1)^r r!(2n+1-r)!\binom nr^{\!2},
 \notag\\
 \alpha_r&=N-1-n+r,\qquad \beta_r=N+n-r.
 \notag
\end{align}
Here $J_i(X)=(-1)^{N-1}P_X'(x_i)$, and similarly for $Y$.
Since $Q_n(x,y)=\mathcal K^{(n)}(x,y)/(y-x)$ has total degree
$2N-2$, double Lagrange interpolation gives
\begin{equation}
 \sum_{i,j}\frac{Q_n(x_i,y_j)}{P_X'(x_i)P_Y'(y_j)}
 =[x^{N-1}y^{N-1}]Q_n(x,y),
 \label{eq:occ_interpolation}
\end{equation}
where brackets denote coefficient extraction.
Every term in parentheses in Eq.~\eqref{eq:occ_root_kernel},
after division by $y-x$, contributes one to this coefficient.
Consequently,
\begin{equation}
 L_n=2g_{p,n}H_n^2
 \binom{N}{n+1}^{\!2}S_n,
 \qquad S_n=\sum_{r=0}^{n}c_{nr}.
 \label{eq:occ_finite_sum}
\end{equation}
The remaining sum follows from
\begin{align}
 \frac{S_n}{(n!)^2(n+1)}
 &=\sum_{r=0}^{n}(-1)^r\binom nr
 \binom{2n+1-r}{n+1}
 \notag\\
 &=[t^{n+1}](1+t)^{2n+1}
 \left(1-\frac{1}{1+t}\right)^n
 \notag\\
 &=[t^{n+1}]t^n(1+t)^{n+1}=n+1.
 \label{eq:occ_sum_evaluation}
\end{align}
Thus $S_n=[(n+1)!]^2$ and
\begin{equation}
 d_np_n=L_n
 =2g_{p,n}H_n^2
 \left[\frac{N!}{(N-1-n)!}\right]^2.
 \label{eq:occ_response_evaluated}
\end{equation}
Finally, inserting
$g_{p,n}H_n^2=(n+1)(N-1-n)!/(N+n+1)!$ gives
\begin{equation}
 \boxed{
 p_n=\frac{(N_1!)^2}{(N_1-1-n)!(N_1+1+n)!},
 \qquad n=0,\ldots,N_1-1.}
 \label{eq:occ_p_exact}
\end{equation}
The result satisfies the particle-number sum rule
\begin{equation}
 \sum_{n=0}^{N_1-1}(2n+2)p_n=N_1.
 \label{eq:occ_sum_rule}
\end{equation}
Indeed, setting $p_{-1}=1$, the summand equals
$(N_1-n)p_{n-1}-(N_1-n-1)p_n$, so the sum telescopes. We plot Eq.~\eqref{eq:occ_p_exact} in Fig.~\ref{fig:cf_occupations_vs_n}.

\section{A Yang pairing state common to both variational families}
\label{app:yang-state}

At $N_\phi=2N_1-1\equiv2q$, the two CF pairing constructions also
contain the electronic state
\begin{equation}
  \Phi_{\mathrm{Yang}}(X,Y)
  =\det_{i,j}\bigl([x_i,y_j]^{N_\phi}\bigr).
  \label{eq:yang-coordinate-state}
\end{equation}
We first derive its two CF representations from the $(111)$
identities, then give its normalized second-quantized form and
describe its antipodal correlations. Throughout, $\propto$ allows a
nonzero coordinate-independent factor, including an overall phase.

\subsection{Derivation from the two $(111)$ identities}

Define the antipodal spinor substitution
\begin{equation}
  \mathcal A(u,v)=(-v^*,u^*),
  \qquad
  \mathcal A Z=(\mathcal A z_1,\ldots,\mathcal A z_{N_1}),
  \label{eq:yang-antipode}
\end{equation}
and let $(\mathsf Dg)_n=(-1)^n g_n$.
Write $F_{n,m;i}(Z)=J_i(Z)\widetilde Y_{n,m;i}(Z)$ for the
polynomial numerators of the projected orbitals. Applying the
antipodal substitution to the JK differential formula gives
\begin{equation}
  F_{n,m;i}(\mathcal A Z)^*
  =(-1)^{\ell_n-m}F_{n,-m;i}(Z),
  \qquad \ell_n=n+\tfrac12.
  \label{eq:yang-orbital-antipode}
\end{equation}
Here $J_i(\mathcal A Z)^*=J_i(Z)$, and the substitution exchanges
the two spinor components and the corresponding derivatives.
The phase in Eq.~\eqref{eq:yang-orbital-antipode} differs from the
$p$-wave contraction phase $(-1)^{m+1/2}$ by $(-1)^{n+1}$.
After absorbing the Jastrow factors into the pairing matrices, this
therefore implies the exact identity
\begin{equation}
  \Phi_{\mathrm{BCS},s}^{\mathrm{ph}}(g;X,\mathcal A Y)
  =(-1)^{N_1}\Phi_{\mathrm{BCS},p}(\mathsf Dg;X,Y).
  \label{eq:yang-antipodal-identity}
\end{equation}

To compare electronic states, define
\begin{align}
  \Phi_s^{\mathrm e}(g)
  &\equiv\mathcal P_\downarrow^{-1}
    \Phi_{\mathrm{BCS},s}^{\mathrm{ph}}(g),\\
  (\mathcal U_\downarrow\Phi)(X,Y)
  &\equiv(\mathcal P_\downarrow\Phi)(X,\mathcal A Y).
  \label{eq:yang-U-definition}
\end{align}
Thus $\mathcal U_\downarrow$ first changes the lower-layer electrons
to holes, and then evaluates each hole coordinate at the antipode
of the corresponding electron coordinate. This is a linear map;
the complex conjugation in Eq.~\eqref{eq:yang-antipode} acts on the
coordinate arguments, not on the coefficients of the state.

At half filling, $\mathcal U_\downarrow$ is an involution. To see
this, label the physical LLL orbitals by
$a=m+q=0,\ldots,2N_1-1$, and let $|R\rangle$ be a normalized
lower-layer Slater determinant with $N_1$ occupied orbitals.
Particle--hole transformation complements $R$, while the antipodal
substitution reverses the orbital labels. Including the determinant
signs, with the holes placed first in the filled-LLL determinant,
gives
\begin{equation}
  \mathcal U_\downarrow|R\rangle
  =(-1)^{N_1}|\rho(R^c)\rangle,
  \qquad \rho(a)=2N_1-1-a.
  \label{eq:yang-U-slater}
\end{equation}
Since reversed complementation squares to the identity,
$\mathcal U_\downarrow^2=1$.
Eq.~\eqref{eq:yang-antipodal-identity} then yields
\begin{align}
  \mathcal U_\downarrow\Phi_s^{\mathrm e}(g)
  &\propto\Phi_{\mathrm{BCS},p}(\mathsf Dg),\\
  \mathcal U_\downarrow\Phi_{\mathrm{BCS},p}(g)
  &\propto\Phi_s^{\mathrm e}(\mathsf Dg).
  \label{eq:yang-family-interchange}
\end{align}

Let $g_p^{(111)}$ and $g_s^{(111)}$ denote the coefficient vectors
reproducing the $(111)$ state in the $p$- and $s$-wave constructions,
respectively. Their components are
\begin{align}
  g_{p,n}^{(111)}
  &=(n+1)
    \frac{(N_1-1-n)!}{(N_1+n+1)!}
    \left[\frac{(2N_1+n)!}{(2N_1)!}\right]^2,
  \label{eq:yang-111-p-coefficients}\\
  g_{s,n}^{(111)}
  &=\delta_{n,N_1-1}.
  \label{eq:yang-111-s-coefficients}
\end{align}
Applying $\mathcal U_\downarrow$ to both known $(111)$ identities
and using
$\mathsf Dg_s^{(111)}=(-1)^{N_1-1}g_s^{(111)}$ gives
\begin{equation}
  \Phi_{\mathrm{BCS},p}\bigl(g_s^{(111)}\bigr)
  \propto\mathcal U_\downarrow\Phi_{111}
  \propto\Phi_s^{\mathrm e}\bigl(\mathsf Dg_p^{(111)}\bigr).
  \label{eq:yang-common-state}
\end{equation}
The middle state is easily identified. The electron--hole $(111)$
kernel is
\begin{equation}
  K_q(x,h)=\frac{2q+1}{4\pi}
    [u(x)u(h)^*+v(x)v(h)^*]^{2q}.
\end{equation}
Since $2q$ is odd,
\begin{equation}
  K_q(x,\mathcal A y)
  =-\frac{2q+1}{4\pi}[x,y]^{2q}.
\end{equation}
Its determinant is proportional to
Eq.~\eqref{eq:yang-coordinate-state}. We have therefore established
\begin{equation}
  \boxed{
  g_{p,n}^{\mathrm{Yang}}=g_{s,n}^{(111)},
  \qquad
  g_{s,n}^{\mathrm{Yang}}=(-1)^n g_{p,n}^{(111)}.}
  \label{eq:yang-pairing-coefficients}
\end{equation}
The $s$-wave state in this statement is understood after the inverse
physical particle--hole transformation. The operation
$\mathcal U_\downarrow$ changes the electronic state, so
Eq.~\eqref{eq:yang-family-interchange} is not a reparametrization of
the same wavefunction for arbitrary coefficients.

\subsection{Normalized second-quantized form}

There are $M=2q+1=2N_1$ physical LLL orbitals in each layer. Let
$c_{\sigma m}^\dagger$ create an electron in layer $\sigma$ and
orbital $m$, and define the orbital-singlet pair operator
\begin{equation}
  B^\dagger
  =\sum_{m=-q}^{q}(-1)^{q-m}
    c_{\uparrow m}^\dagger c_{\downarrow,-m}^\dagger.
  \label{eq:yang-pair-operator}
\end{equation}
The phases are the Clebsch--Gordan phases coupling two angular
momenta $q$ to total orbital angular momentum $L=0$.
With the orbital convention
\begin{equation}
  \varphi_m(x)=(-1)^{q-m}
    \sqrt{\frac{M}{4\pi}\binom{2q}{q-m}}
    u(x)^{q+m}v(x)^{q-m},
\end{equation}
the corresponding coordinate pair kernel follows from the binomial
theorem:
\begin{align}
  K_B(x,y)
  &\equiv\sum_{m=-q}^{q}(-1)^{q-m}
    \varphi_m(x)\varphi_{-m}(y)
    \nonumber\\
  &=-\frac{M}{4\pi}[x,y]^{2q}.
  \label{eq:yang-pair-kernel}
\end{align}
The minus sign is an overall pair phase and has no physical effect.
Antisymmetrizing $N_1$ copies of this pair orbital produces
$\det_{i,j}K_B(x_i,y_j)$, which is proportional to
$\Phi_{\mathrm{Yang}}$.

For the normalization, introduce the orthonormal lower-layer modes
and canonical pair creators
\begin{equation}
  d_m^\dagger=(-1)^{q-m}c_{\downarrow,-m}^\dagger,
  \qquad b_m^\dagger=c_{\uparrow m}^\dagger d_m^\dagger.
\end{equation}
Different $b_m^\dagger$ commute, while $(b_m^\dagger)^2=0$.
Consequently,
\begin{equation}
  (B^\dagger)^{N_1}|0\rangle
  =N_1!\sum_{\substack{R\subset\{-q,\ldots,q\}\\|R|=N_1}}
    \prod_{m\in R}b_m^\dagger|0\rangle.
  \label{eq:yang-subset-expansion}
\end{equation}
The $\binom{2N_1}{N_1}$ subset states on the right are orthonormal.
The normalized second-quantized state is therefore
\begin{equation}
  \boxed{
  |\Psi_{\mathrm{Yang}}\rangle
  =\frac{(B^\dagger)^{N_1}}
    {N_1!\sqrt{\binom{2N_1}{N_1}}}|0\rangle
  =\frac{(B^\dagger)^{N_1}}{\sqrt{(2N_1)!}}|0\rangle.}
  \label{eq:yang-normalized-fock-state}
\end{equation}
It is an equal-weight superposition of all choices of $N_1$
occupied canonical pair modes out of $2N_1$, which identifies it
as a Yang pairing state~\cite{Yang1962concept,Carlen2016entropy}.
Equivalently, it is the component with $N_1$ electrons in each layer
of the BCS product $\prod_m(1+b_m^\dagger)|0\rangle$.

For completeness, use the Fock ordering with all upper-layer
creators preceding all lower-layer creators, and increasing $m$
within each layer. Moving all upper-layer creators to the left of
the lower-layer creators contributes $(-1)^{N_1(N_1-1)/2}$;
the determinant below includes antisymmetrization within each layer.
With the product angular measure and no factorial prefactors,
the normalized coordinate wavefunction is then
\begin{align}
  \Psi_{\mathrm{Yang}}(X,Y)
  &=\frac{(-1)^{N_1(N_1-1)/2}}{\sqrt{(2N_1)!}}
    \det_{i,j}K_B(x_i,y_j)
    \nonumber\\
  &=\frac{(-1)^{N_1(N_1+1)/2}}{\sqrt{(2N_1)!}}
    \left(\frac{M}{4\pi}\right)^{N_1}
    \Phi_{\mathrm{Yang}}(X,Y).
  \label{eq:yang-normalized-coordinate-state}
\end{align}
Thus
\begin{equation}
  \|\Phi_{\mathrm{Yang}}\|^2
  =\left(\frac{4\pi}{M}\right)^{2N_1}(2N_1)!.
  \label{eq:yang-coordinate-norm}
\end{equation}
The displayed overall phase may be suppressed when comparing
physical states. Here ``singlet'' refers to orbital angular
momentum; it does not denote a singlet of the layer pseudospin.

\subsection{Pair density matrix and antipodal correlations}

The equal-weight expansion also gives the interlayer two-particle
density matrix directly. In the canonical orbital basis, define
\begin{equation}
  \Gamma_{mn;kl}
  =\langle c_{\uparrow m}^\dagger d_n^\dagger
    d_l c_{\uparrow k}\rangle,
  \qquad \operatorname{Tr}\Gamma=N_1^2.
\end{equation}
Counting subsets in Eq.~\eqref{eq:yang-subset-expansion} yields
\begin{align}
  \Gamma_{mn;kl}
  &=\alpha_0\delta_{mk}\delta_{nl}
    +\alpha_1\delta_{mn}\delta_{kl},
  \label{eq:yang-two-particle-matrix}\\
  \alpha_0&=\frac{N_1(N_1-1)}{M(M-1)},
  \qquad
  \alpha_1=\frac{N_1(M-N_1)}{M(M-1)}.
\end{align}
The first coefficient is the probability that two distinct
canonical pair modes are occupied. The second is the matrix
element for transferring a pair between two distinct modes.
The normalized eigenvector $M^{-1/2}\sum_m|m,m\rangle$ has
eigenvalue
\begin{equation}
  \lambda_{\mathrm{pair}}
  =\alpha_0+M\alpha_1
  =\frac{N_1+1}{2},
  \label{eq:yang-pair-eigenvalue}
\end{equation}
while every orthogonal eigenvector has eigenvalue $\alpha_0$.
The eigenvalue in Eq.~\eqref{eq:yang-pair-eigenvalue} is extensive,
but its physical interpretation depends on the spatial form of the
occupied pair orbital, $K_B(x,y)/\sqrt M$.

For two points separated by an angle $\gamma$,
$|[x,y]|^2=\sin^2(\gamma/2)$. Hence
\begin{equation}
  |K_B(x,y)|^2
  =\left(\frac{M}{4\pi}\right)^2
    \sin^{4q}\!\left(\frac{\gamma}{2}\right).
  \label{eq:yang-antipodal-kernel}
\end{equation}
The pair amplitude is maximal at antipodal points.
Writing $\delta=\pi-\gamma$ and $s=R\delta$, where
$R^2=q\ell_B^2$, gives near the antipode
\begin{equation}
  \sin^{4q}\!\left(\frac{\gamma}{2}\right)
  =\cos^{4q}\!\left(\frac{\delta}{2}\right)
  \simeq \exp\!\left(-\frac{q\delta^2}{2}\right)
  =\exp\!\left(-\frac{s^2}{2\ell_B^2}\right).
  \label{eq:yang-antipodal-width}
\end{equation}
Thus the width about the antipode is of order $\ell_B$, whereas
the distance between the paired electrons grows with $R$.

The actual many-body density correlation follows from
Eq.~\eqref{eq:yang-two-particle-matrix}. Let $\rho_\sigma(x)$ be
the density per unit solid angle, so
$\langle\rho_\sigma\rangle=N_1/(4\pi)$. Then
\begin{equation}
  \langle\rho_\uparrow(x)\rho_\downarrow(y)\rangle
  =\alpha_0\left(\frac{M}{4\pi}\right)^2
    +\alpha_1|K_B(x,y)|^2.
\end{equation}
The normalized interlayer correlation function is therefore
\begin{equation}
  \boxed{
  g_{\uparrow\downarrow}^{\mathrm{Yang}}(\gamma)
  =\frac{2N_1-2}{2N_1-1}
    +\frac{2N_1}{2N_1-1}
      \sin^{4N_1-2}\!\left(\frac{\gamma}{2}\right).}
  \label{eq:yang-density-correlation}
\end{equation}
In particular, $g_{\uparrow\downarrow}^{\mathrm{Yang}}(\pi)=2$.
In the planar thermodynamic limit at fixed physical separation
$r=R\gamma$, however,
\begin{equation}
  g_{\uparrow\downarrow}^{\mathrm{Yang}}(r)\longrightarrow1.
\end{equation}
The enhancement moves to infinite separation. Accordingly, the
extensive pair-density eigenvalue does not establish a local
interlayer pair condensate in this limit. The additional common
state should be interpreted as an exact finite-size intersection
of the two variational families whose pairing correlations depend
on the global antipodal geometry.

\section{A counterexample to equivalence of the two variational families}
\label{app:sp-counterexample}

The two paired constructions reproduce the same electronic state at
special choices of their coefficients, including the $(111)$ state.
Here we show that the two variational families contain different sets
of electronic states. In particular, at $N_1=4$ there is an $s$-wave
state which, after transforming the lower-layer holes back to
electrons, cannot be represented by the $p$-wave ansatz for any choice
of its coefficients. Define
\begin{equation}
  \Phi_s^{\mathrm e}(g;X,Y)
  \equiv \mathcal P_\downarrow^{-1}
  \Phi_{\mathrm{BCS},s}^{\mathrm{ph}}(g;X,H),
  \label{eq:sp-counter-electron-state}
\end{equation}
where $\mathcal P_\downarrow$ is the particle--hole change of basis
with the holes placed first in the filled-LLL determinant.
Proportionality below always means equality up to a nonzero,
coordinate-independent factor.

\subsection{Layer-exchange symmetry of the $p$-wave family}

Introduce the polynomial numerators
$F_{n,m;i}(Z)=J_i(Z)\widetilde Y_{n,m;i}(Z)$.
The Jastrow factors can be absorbed into the pairing matrix, giving
\begin{align}
  \Phi_{\mathrm{BCS},p}(g;X,Y)&=\det K_p(g;X,Y),
  \label{eq:sp-counter-p-det}\\
  [K_p(g;X,Y)]_{ij}
  &=\sum_{n=0}^{N_1-1}\sum_{m=-\ell_n}^{\ell_n}
    g_n(-1)^{m+1/2}
    F_{n,m;i}(X)F_{n,-m;j}(Y),
  \qquad \ell_n=n+\tfrac12.
  \label{eq:sp-counter-p-kernel}
\end{align}
Changing $m\mapsto-m$ and using
$(-1)^{-m+1/2}=-(-1)^{m+1/2}$ yields
\begin{equation}
  K_p(g;Y,X)=-K_p(g;X,Y)^T.
\end{equation}
Consequently, every $p$-wave state obeys
\begin{equation}
  \Phi_{\mathrm{BCS},p}(g;Y,X)
  =(-1)^{N_1}\Phi_{\mathrm{BCS},p}(g;X,Y),
  \label{eq:sp-counter-p-parity}
\end{equation}
including for complex $g_n$. For $N_1=4$, this is even parity under
exchange of the two coordinate sets.

To test this condition, set $N_\phi=7$ and work in the affine patch
$(u_i,v_i)=(1,z_i)$; the homogeneous spinor factors can be restored
uniquely. For a four-element subset $R\subset\{0,\ldots,7\}$,
ordered increasingly, define the physical monomial Slater determinant
\begin{equation}
  M_R(Z)=\det_{i,r\in R}z_i^r.
  \label{eq:sp-counter-monomial-slater}
\end{equation}
These determinants form a basis of the one-layer four-electron LLL
space. Thus, if
\begin{equation}
  \Phi(X,Y)=\sum_{I,K}C_{I,K}M_I(X)M_K(Y),
  \label{eq:sp-counter-electron-expansion}
\end{equation}
membership in the $p$-wave family requires
$C_{I,K}=C_{K,I}$ for every pair of subsets.

\subsection{An explicit $s$-wave state}

Choose the allowed coefficient vector
\begin{equation}
  g_{s,n}=\delta_{n,2},\qquad n=0,1,2,3.
  \label{eq:sp-counter-choice}
\end{equation}
The occupied CF shell has $\ell=5/2$ and six orbitals. Label these
auxiliary orbitals by $a=5/2-m=0,\ldots,5$, keeping $a$ distinct from
the physical LLL exponent $r$. Write
\begin{equation}
  F_{2,5/2-a;i}(Z)
  =\kappa\sqrt{a!(5-a)!}\,f_{ia}(Z),
  \qquad \kappa=\frac{8!}{10!\sqrt{2\pi}}.
  \label{eq:sp-counter-core-normalization}
\end{equation}
The JK differential formula gives the following explicit polynomial:
\begin{equation}
  f_{ia}(Z)=\left.
  \frac{(-1)^a}{2}
  \sum_{s=0}^{2}(-1)^s\binom{2}{s}\binom{3}{a-s}
  u_i^{3-a+s}v_i^{a-s}
  \partial_{u_i}^{s}\partial_{v_i}^{2-s}J_i(Z)
  \right|_{(u_j,v_j)=(1,z_j)}.
  \label{eq:sp-counter-core-formula}
\end{equation}
Terms with vanishing binomial coefficients are omitted before
evaluating spinor powers. Each $f_{ia}$ has total affine degree
$a+1$.

For a four-element auxiliary subset $A\subset\{0,\ldots,5\}$, define
\begin{equation}
  f_A(Z)=\det_{i,a\in A}f_{ia}(Z),
  \qquad \eta_A=\prod_{a\in A}a!(5-a)!.
  \label{eq:sp-counter-auxiliary-slater}
\end{equation}
Cauchy--Binet then gives the full electron--hole wavefunction,
\begin{equation}
  \Phi_{\mathrm{BCS},s}^{\mathrm{ph}}(\delta_{n,2};X,H)
  =\kappa^8\sum_{|A|=4}\eta_A f_A(X)f_A(H)^*.
  \label{eq:sp-counter-cauchy-binet}
\end{equation}
The factors $\eta_A$ retain all relative auxiliary-orbital
normalizations.

Only two of these minors will be needed. Let
$A_{\mathrm h}=\{0,1,2,3\}$ and
$A_{\mathrm l}=\{2,3,4,5\}$.
Since $f_A$ has total degree $4+\sum_{a\in A}a$,
$A_{\mathrm h}$ is the unique subset contributing degree $10$, and
$A_{\mathrm l}$ is the unique subset contributing degree $18$.

The required minor can be evaluated by a small determinant. Let
$e_k=e_k(z_1,z_2,z_3,z_4)$ be the elementary symmetric polynomials.
Using
$z_i^4=e_1z_i^3-e_2z_i^2+e_3z_i-e_4$ in
Eq.~\eqref{eq:sp-counter-core-formula} gives
\begin{equation}
  (f_{i0},f_{i1},f_{i2},f_{i3})
  =(1,z_i,z_i^2,z_i^3)
  \begin{pmatrix}
    e_1 & -2e_2 & 3e_3 & -4e_4\\
    -4 & e_1 & 2e_2 & -5e_3\\
    0 & 8 & -5e_1 & 2e_2\\
    0 & 0 & -4 & 3e_1
  \end{pmatrix}.
  \label{eq:sp-counter-small-matrix}
\end{equation}
Taking its determinant yields
\begin{align}
  f_{A_{\mathrm h}}(Z)
  &=M_{0123}(Z)
    \bigl(-15e_1^4+80e_1^2e_2-64e_2^2
          -128e_1e_3+512e_4\bigr)
    \nonumber\\
  &=-15M_{0127}+35M_{0136}-14M_{0145}
    -77M_{0235}+385M_{1234}.
  \label{eq:sp-counter-high-expansion}
\end{align}
Here $M_{0127}$ denotes $M_{\{0,1,2,7\}}$, and similarly for the
other subscripts. The second equality is an ordinary polynomial
identity, obtained by expanding the monomial determinants.

The opposite minor follows from the identity
\begin{equation}
  f_{i,5-a}(Z)
  =z_i^4\!\prod_{j\ne i}z_j\,f_{ia}(Z^{-1}),
  \qquad Z^{-1}=(z_1^{-1},\ldots,z_4^{-1}).
\end{equation}
Reversing four determinant columns has positive sign, so
\begin{equation}
  f_{A_{\mathrm l}}(Z)
  =\left(\prod_{i=1}^{4}z_i^7\right)f_{A_{\mathrm h}}(Z^{-1}).
  \label{eq:sp-counter-reflection}
\end{equation}
The identities involving $Z^{-1}$ initially hold for nonzero $z_i$
and extend to all coordinates as polynomial identities.
Thus the coefficient of $M_{\rho(R)}$ in $f_{A_{\mathrm l}}$ equals
the coefficient of $M_R$ in $f_{A_{\mathrm h}}$, where
$\rho(R)=\{7-r:r\in R\}$ is sorted increasingly.
Both relevant auxiliary weights are equal:
\begin{equation}
  \eta_{A_{\mathrm h}}=\eta_{A_{\mathrm l}}
  =120\cdot24\cdot12\cdot12\equiv\eta_\star.
  \label{eq:sp-counter-equal-weights}
\end{equation}

\subsection{Transforming back to electrons}

Particle--hole transformation must be applied in an orthonormal
physical LLL basis. The binomial normalization factors therefore
matter when using the monomial determinants $M_R$.
At $N_\phi=7$, choose
\begin{equation}
  \varphi_r(u,v)=(-1)^r
  \sqrt{\frac{2}{\pi}\binom{7}{r}}\,u^{7-r}v^r,
  \qquad r=0,\ldots,7.
\end{equation}
Let $V_R$ be the corresponding Slater determinant and define
\begin{equation}
  w_R=\prod_{r\in R}\binom{7}{r},
  \qquad
  V_R=\left(\frac{2}{\pi}\right)^2
  (-1)^{\sum_{r\in R}r}\sqrt{w_R}\,M_R
  \label{eq:sp-counter-physical-metric}
\end{equation}
in the affine patch. With the stated particle--hole convention,
\begin{equation}
  \mathcal P_\downarrow^{-1}V_R(H)^*
  =\epsilon(R)V_{R^c}(Y),
  \qquad \epsilon(R)=(-1)^{\sum_{r\in R}r-6},
\end{equation}
where the complement is taken in $\{0,\ldots,7\}$.
Since the sum of all eight exponents is $28$, the orbital phase
factors cancel between $R$ and $R^c$. Also,
$w_Rw_{R^c}=5145^2$. Hence
\begin{equation}
  \mathcal P_\downarrow^{-1}M_R(H)^*
  =\epsilon(R)\frac{5145}{w_R}M_{R^c}(Y).
  \label{eq:sp-counter-monomial-ph}
\end{equation}
Dividing all $V_R$ by $\sqrt{4!}$ gives normalized many-electron
Slaters and leaves this relation unchanged.

Now choose the physical subsets
\begin{equation}
  I=\{0,1,4,5\},\qquad K=\{3,4,5,6\},
\end{equation}
with
$K^c=\{0,1,2,7\}$ and $I^c=\{2,3,6,7\}$.
Their total degrees are $\sum I=10$ and $\sum K=18$.
Therefore, in the complete sum
\eqref{eq:sp-counter-cauchy-binet}, the coefficient $C_{I,K}$
receives a contribution only from $A_{\mathrm h}$, whereas
$C_{K,I}$ receives a contribution only from $A_{\mathrm l}$.
Equations~\eqref{eq:sp-counter-high-expansion}
and~\eqref{eq:sp-counter-reflection} give the four required
Slater coefficients:
\begin{equation}
  \begin{array}{c|cc}
    \text{auxiliary minor} & \text{physical subset} &
       \text{coefficient}\\ \hline
    f_{A_{\mathrm h}} & I & -14\\
    f_{A_{\mathrm h}} & K^c & -15\\
    f_{A_{\mathrm l}} & K & 385\\
    f_{A_{\mathrm l}} & I^c & -14
  \end{array}
\end{equation}
The particle--hole factors are
\begin{equation}
  \epsilon(K^c)=\epsilon(I^c)=1,
  \qquad w_{K^c}=147,\qquad w_{I^c}=5145.
\end{equation}
It follows, without numerical approximation, that
\begin{align}
  C_{I,K}
  &=\kappa^8\eta_\star(-14)(-15)\frac{5145}{147}
    =7350\,\kappa^8\eta_\star,\\
  C_{K,I}
  &=\kappa^8\eta_\star(385)(-14)\frac{5145}{5145}
    =-5390\,\kappa^8\eta_\star.
\end{align}
In particular, the state is nonzero and
\begin{equation}
  \boxed{\frac{C_{K,I}}{C_{I,K}}=-\frac{11}{15}\ne1.}
  \label{eq:sp-counter-obstruction}
\end{equation}
Using normalized physical Slaters instead of $M_I M_K$ multiplies
both coefficients by the same factor, so the ratio is unchanged.

This violates the necessary layer-exchange symmetry
\eqref{eq:sp-counter-p-parity}. Consequently,
\begin{equation}
  \mathcal P_\downarrow^{-1}
  \Phi_{\mathrm{BCS},s}^{\mathrm{ph}}(\delta_{n,2})
  \not\propto\Phi_{\mathrm{BCS},p}(g_p)
  \qquad\text{for every }g_p\in\mathbb C^4.
  \label{eq:sp-counter-conclusion}
\end{equation}
Thus no map of pairing coefficients, even a nonlinear one, can
identify the same electronic states throughout the two variational
families. The nonzero difference $C_{I,K}-C_{K,I}$ is continuous in
the $s$-wave coefficients, so the obstruction also survives
sufficiently small positive admixtures of the other shells.
This conclusion concerns the finite-size ans\"atze as defined, before
any additional layer symmetrization. It does not exclude their
special common states, or an invertible transformation that changes
the physical state while mapping one family to the other, such as
the particle--hole transformation followed by antipodal coordinate
substitution.

\end{appendix}

\end{document}